\documentclass[preprint]{aastex}
\usepackage{psfig}

\newcommand{\Ha}{H$\alpha$}

\newcommand{\HII}{\ion{H}{2}}

\shorttitle{X-RAY PROPERTIES OF LOW-LUMINOSITY AGNs}
\shortauthors{HALDERSON ET AL.}

\begin{document}

\title{The Soft X-ray Properties of Nearby Low-Luminosity Active
       Galactic Nuclei and their Contribution to the Cosmic X-ray Background}

\author{Eve L.\ Halderson, Edward C.\ Moran\altaffilmark{1},
        Alexei V.\ Filippenko}

\affil{Department of Astronomy, University of California,
                 Berkeley, CA 94720-3411}

\altaffiltext{1}{Chandra Fellow.}

\author{Luis C.\ Ho}

\affil{The Observatories of the Carnegie Institution of Washington,
                 813 Santa Barbara St., Pasadena, CA 91101}

\begin{abstract}

We have examined {\sl ROSAT\/} soft X-ray observations of a complete,
distance-limited sample of Seyfert and LINER galaxies.  X-ray data are
available for 46 out of 60 such objects which lie within a hemisphere of
radius 18 Mpc.  We have constructed radial profiles of the nuclear sources
in order to characterize their spatial extent and, in some cases, to help
constrain the amount of flux associated with a nuclear point source.  PSPC
data from {\sl ROSAT\/} have been used to explore the spectral characteristics
of the objects with sufficient numbers of detected counts.  Based on the
typical spectral parameters of these sources, we have estimated the
luminosities of the weaker sources in the sample.  We then explore the
relationship between the soft X-ray and \Ha\ luminosities of the observed
objects; these quantities are correlated for higher-luminosity AGNs.  We
find a weak correlation at low luminosities as well, and have used this
relationship to predict $L_{\rm X}$ for the 14 objects in our sample that
lack X-ray data.  Using the results of the spatial and spectral analyses,
we have compared the X-ray properties of Seyferts and LINERs, finding no
striking differences between the two classes of objects.  However, both 
types of objects often exhibit significant amounts of extended emission, 
which could minimize the appearance of differences in their nuclear 
properties.  The soft X-ray characteristics of the type 1 and type 2 active 
galaxies in the sample are also discussed.  We then compute the local X-ray 
volume emissivity of low-luminosity Seyferts and LINERs and investigate 
their contribution to the cosmic X-ray background.  The 0.5--2.0 keV volume
emissivity of $2.2 \times 10^{38}$ ergs~s$^{-1}$~Mpc$^{-3}$ we obtain
for our sample suggests that low-luminosity AGNs produce at least 9\%
of the soft X-ray background.

\end{abstract}

\keywords{galaxies:~active --- galaxies: nuclei --- galaxies: Seyfert --- 
X-rays: galaxies}

\section{Introduction}

The bolometric luminosities of active galactic nuclei (AGNs) span more
than eight orders of magnitude from the most powerful quasars to the
weakest known Seyfert galaxies.  Despite their luminosity differences,
however, quasars and Seyferts possess many of the same characteristics.
In particular, their rest-frame optical spectra are remarkably similar,
exhibiting a luminosity-independent relationship between the emission-line
and nonstellar continuum strengths (Yee 1980; Shuder 1981; Ho \& Peng 2001).  
Thus, it has been suspected for some time that nearby, low-luminosity 
Seyfert nuclei and bright, distant quasars are powered by the same 
fundamental mechanism---accretion of matter onto a supermassive black hole.  

The questions facing us now are whether there is a lower luminosity limit 
below which this mechanism does not operate, and how prevalent this type of
nuclear activity is among galaxies today.  In this context, the nature
of the processes that power low-ionization nuclear emission-line regions
(LINERs; Heckman 1980), which are found in a significant fraction of
all nearby galaxies (Ho, Filippenko, \& Sargent 1997b), is of particular
importance.  Compared to Seyferts, LINERs display a lower degree of
ionization in their emission-line spectra that could result
from photoionization by a dilute nonstellar continuum associated with
an AGN (Halpern \& Steiner 1983; Ferland \& Netzer 1983; Ho, Filippenko, 
\& Sargent 1993).  However, other astrophysical processes are capable of 
producing LINER-like nebular spectra, including shocks from supernova 
explosions (Heckman 1980; Dopita \& Sutherland 1995) and photoionization 
by extremely hot stars in compact starbursts (Filippenko \& Terlevich 1992; 
Shields 1992; Barth \& Shields 2000).  It may well be that LINERs are a 
heterogeneous class whose optical emission lines are powered in some cases 
by accretion and in others by stellar processes (e.g., Filippenko 1996).  
Determination of the fraction of LINERs that contain genuine active nuclei 
is needed to refine estimates of the incidence of supermassive black holes 
in galactic nuclei, an issue which is central to our understanding of galaxy 
formation and evolution.  It is also vital for establishing the shape of the 
faint end of the AGN luminosity function, and in turn, the total contribution 
of AGNs to the extragalactic X-ray background (Comastri et al.\ 1995).

Observations of the luminosity, spectrum, and spatial extent of X-ray
emission in galaxies are particularly useful for addressing the nature
of their nuclear activity.  For example, the soft X-ray emission of
star-forming galaxies often includes thermal components that are extended
on kiloparsec scales, whereas AGNs are typically associated with unresolved,
nonthermal X-ray sources.  Detection of a point-like X-ray source in
the nucleus of M81 using the {\sl Einstein\/} High Resolution Imager
(Elvis \& van Speybroeck 1982) played a key role in establishing the
presence of a weak AGN in this nearby object.  As the first AGN to be
identified with a soft X-ray luminosity of $\sim 10^{40}$ ergs~s$^{-1}$,
this discovery extended the AGN luminosity function two orders of magnitude
below the previously achieved limit.  More recently, Koratkar et al.\
(1995) have investigated five other low-luminosity AGNs (LLAGNs) with
{\sl ROSAT}, finding that they resemble classical broad-line Seyfert
galaxies in terms of the spatial extent and spectral characteristics of
their nuclear X-ray emission.  We might expect, therefore, that a careful
comparison of the X-ray properties of LINERs and low-luminosity Seyfert
galaxies would allow us to test for a common origin of the nuclear
activity in the two types of objects.

Soft X-ray data are also useful for the investigation of AGN unification.
It has been suggested that radio-quiet AGNs with and without broad optical
emission lines (type~1 and type~2, respectively) are intrinsically similar
objects that appear to be different because of the presence or absence of
dense circumnuclear material along our line of sight (Antonucci 1993).  The
material thought to be obscuring the broad emission-line region and the
optical/UV continuum source in type~2 AGNs should absorb a significant
fraction of their soft X-ray fluxes as well (Mulchaey et al.\ 1993).  Thus,
a soft X-ray survey of a well-defined sample of AGNs containing both type~1
and type~2 objects might provide insight as to whether or not the unified
model applies to the majority of AGNs.

To explore these issues, we have assembled a relatively unbiased, 
distance-limited sample of low-luminosity Seyferts and LINERs.  The 
majority have been observed by {\sl ROSAT\/} in the soft (0.1--2.4 keV) 
X-ray band.  In the next two sections, we describe the sample of objects 
and the available data.  In \S~4, we examine the spatial properties of 
the nuclear X-ray sources and present an analysis of their soft X-ray 
fluxes and luminosities.  A comparison of our results for the various 
subclasses of objects represented in our sample is discussed in \S~5, 
along with our investigation of the LLAGN contribution to the soft X-ray 
background.

\section{A Distance-Limited LLAGN Sample}

Between 1984 and 1990, Ho, Filippenko, \& Sargent (1995, 1997a, 1997b; Ho 
et al. 1997c; see also Filippenko \& Sargent 1985) carried out an optical 
spectroscopic survey of nearby galactic nuclei in the northern sky.  
Galaxies with total apparent blue magnitudes $B_{\rm T} < 12.5$ were 
included in the survey, which is nearly complete for $B_{\rm T} < 12.0$ mag.  
Most (418) of the 486 objects observed have emission-line spectra indicative 
of some sort of nuclear activity (Ho et al.\ 1997b): 49\% of the emission-line 
nuclei are classified as star-forming (\HII ) galaxies, 23\% are LINERs, 
16\% have ``transition'' spectra, which may represent a combination of \HII\ 
and LINER characteristics, and the remainder (12\%) are Seyfert galaxies of 
various types.

Within a distance of 18~Mpc, 60 galaxies from the Ho et al.\ survey are
classified as Seyferts (22) or LINERs (38).  The names, spectroscopic
classifications, distances,\footnote{Distances from Ho et al. (1997a) 
assume $H_{0} = 75$ km~s$^{-1}$~Mpc$^{-1}$, which we adopt throughout this
paper.  The distributions of distances for Seyferts and LINERs and for
type 1 and type 2 objects are extremely similar, so we may compare different
classes of objects without concern for systematic differences in distances.} 
and \Ha\ luminosities of these objects (from Ho et al.\ 1997a) are 
listed in Table~1.  Only one galaxy, NGC 1068, has an \Ha\ luminosity 
substantially greater than $10^{40}$ ergs~s$^{-1}$; as Ho et al.\ (1997a) 
have discussed, this limit is useful for distinguishing high- and 
low-luminosity AGNs.  Thus, since the Ho et al.\ survey is expected to be 
complete for nucleated galaxies out to 18 Mpc, the 60 Seyferts and LINERs 
in our list constitute a large, relatively unbiased sample of low-luminosity 
AGNs and AGN candidates.\footnote{The nature of the nuclear activity in the 
transition objects is still under investigation, so we have omitted them from 
the present study.  See Ho \& Ulvestad (2001, Appendix A) for a discussion 
of issues concerning completeness and selection biases in AGN samples,
including the one studied here.}

\section{Soft X-ray Observations}

The majority of the objects in our sample have been observed with the
{\sl ROSAT\/} Observatory, using the High Resolution Imager (HRI) and
the Position Sensitive Proportional Counter (PSPC) instruments on board.
Both instruments are sensitive in an energy range of approximately
0.1--2.4 keV.  The HRI provides an angular resolution (half-energy width)
of about $5''$ on-axis, but has very little spectral sensitivity.  Its
field of view is $38'$ square, most of which is usable.  The PSPC has
lower angular resolution ($25''$ at 1 keV), but it has modest spectral
sensitivity, with an energy resolution $E/\Delta E = 0.43 (E/0.93)^{-0.5}$.
The full PSPC field of view covers a radius of $1^{\circ}$; however,
vignetting and broadening of the point-response function degrade the
image quality at off-axis angles greater than $\sim 10'$ (see the
{\sl ROSAT\/} Mission Description, Appendix F).

{\sl ROSAT\/} images of 44 of the 60 objects in our sample are available.
We acquired these data through HEASARC at NASA's Goddard Space Flight Center.
A total of 21 galaxies were observed with both the PSPC and HRI instruments.  
Another 15 objects were observed only with the PSPC, and 8 were observed with 
the HRI only.  Two galaxies that were not observed by {\sl ROSAT}, NGC 4698 
and NGC 4762, were detected with the {\sl Einstein\/} Imaging Proportional 
Counter (IPC) in the 0.2--3.5 keV band (Fabbiano, Kim, \& Trinchieri 1992).  
Thus, the majority of our sample (46 galaxies, or 77\%) has some type of soft 
X-ray data available.

For most of the galaxies, no more than one observation per {\sl ROSAT\/}
instrument was made.  A handful of objects, however, were observed
multiple times, often for monitoring purposes.  We combined multiple
images of a galaxy when its position in the field was the same.
Otherwise, a single image providing both the longest individual exposure
time and smallest off-axis angle was analyzed, except when the galaxy
was partially obscured by one of the PSPC window support ribs.  In such
cases, an alternate image providing an unobstructed view of the object
was used.

Exposure times range from 5~ks to 115~ks for the 29 HRI observations, and
in each the galaxy is located less than $6'$ off-axis.  Source counts were
extracted within a circular region large enough to include all of the flux
arising near the nucleus ($20''$--$40''$ in radius for point-like sources).
A background count rate was determined in a concentric annulus.  Contaminating 
sources in the background region were excluded, and the size of the 
annulus was chosen to ensure that at least 100 background counts were 
collected.  All but six HRI sources were detected; 3~$\sigma$ upper limits 
were computed for the non-detections.  The majority of the 36 galaxies imaged 
with the PSPC are located within 15$'$ of the field center, although we
analyzed data for some objects observed as much as 45$'$ off-axis.  PSPC 
exposure times range from 1~ks to 95~ks for our galaxies.  Again, nuclear 
counts were extracted within a circular region large enough to include all 
of the associated flux, and the background was measured in a concentric 
annulus.  As with the HRI, we ensured that PSPC backgrounds were well 
sampled and free of resolved sources.  Net count rates for off-axis PSPC 
sources were corrected for vignetting.  We computed upper limits for the 
four PSPC sources that were not detected.  For each of the 46 galaxies 
observed, Table~2 lists the instrumentation used along with the associated 
exposure time, background-subtracted nuclear count rate, and off-axis angle.

\section{Analysis and Results}

The {\sl ROSAT\/} data provide information about the luminosities and the
spatial and spectral characteristics of our galaxies in the soft X-ray band.
These properties are useful for comparing different subclasses of objects
and for quantifying the LLAGN contribution to the X-ray background.  We
begin with an investigation of the spatial profiles of the nuclear X-ray
sources.  We then present spectral analysis and our determination of the
nuclear X-ray fluxes for the entire sample, along with a comparison to
previously published results.

\subsection{Spatial Analysis}

Our investigation of the soft X-ray spatial characteristics of the galaxies
in the LLAGN sample involves a comparison of their azimuthally averaged radial
count distributions (radial profiles) to theoretical point-spread functions
(PSFs) for the {\sl ROSAT\/} instruments.  Since our primary objective is to
study the X-ray emission associated with the galaxy nuclei, we have omitted
any discrete extranuclear sources from our analysis.  We used the ``centroid''
procedure in the {\tt XIMAGE} software to determine the position of the
central X-ray source in each object.  To confirm that the X-ray emission is
indeed associated with the nucleus, we compared the centroid with the measured
optical position reported by Cotton, Condon, \& Arbizzani (1999), which is 
accurate to $\sim 2''$.  The optical/X-ray position offsets $\Delta_{\rm OX}$ 
are listed in the last column of Table 2 for both the PSPC and HRI data.  
Most of the X-ray centroids are within a few arcseconds of the optical 
nucleus, and all are within $20''$.  Values of $\Delta_{\rm OX}$ obtained 
with the PSPC, despite the lower angular resolution of that instrument, are 
comparable to those obtained with the HRI.  Given the expected uncertainty 
in the X-ray source positions (see Roberts \& Warwick 2000), the small 
optical/X-ray offsets suggest that the central X-ray source in each galaxy 
is associated with the optical nucleus.

Because of the broadening of the {\sl ROSAT\/} PSF with off-axis angle and
other limitations imposed by the detectors, we have restricted spatial
analysis to sufficiently bright sources ($> 50$ counts net), and in the 
case of the PSPC, to sources detected within $18'$ of the optical axis,
where the window support ring is located.  A total of 31 different objects
(22 observed with HRI and 28 observed with PSPC) satisfy these criteria.  We 
used the FTOOLS programs {\tt extrpsf} and {\tt calcrpsf} to generate a 
radial profile for the nuclear sources and a theoretical PSF corresponding 
to their location on the detector.   The accuracy of the radial profile is 
somewhat sensitive to the adopted position of the X-ray source---off-center 
positions diminish the peak of the profile and broaden it as well.  
Therefore, for a few point-like sources, we constructed radial profiles 
using both the measured centroid and adjacent pixels as the source position.  
This verified that the centroids yielded by {\tt XIMAGE} are accurate.

Examples of radial profiles covering the range of characteristics present
in the LLAGN sample (point-like, moderately resolved, and very extended)
are displayed in Figure~1.  The theoretical PSF is plotted with each
profile.  Because of variations in the shape of the PSF, the quality of
the detections, and distances of the sources, it is difficult to provide
a uniform measure of source extent for the sample.  To approximate how
point-like each source is, we have computed the ratio of the maximum number
of counts expected for a point source (from the theoretical PSF) to the
total counts observed.  For the measurement, we scaled the PSF such that
it (1) peaks at a surface brightness 1~$\sigma$ higher than that of the
innermost bin of the radial profile, and (2) has a zero-point consistent
with the measured background in the image.  The derived ratio is therefore
a conservative upper limit on the fraction of the observed counts that can
be associated with a point source.  The results, for both the PSPC and HRI
profiles, are listed in Table 3.  As the Table shows, only a handful of
the sources can be considered point-like; the central X-ray emission in
most of the galaxies appears to have a significant extended component, even
at the resolution of the PSPC.  This comes as a mild surprise, given that
our sample consists of AGNs.  When galaxies have been observed with both
the PSPC and HRI, the HRI point-source fractions are almost always lower,
indicating that the source extents are real.

In principle, we could use the derived point-source fractions---particularly
those obtained with the HRI---to improve our estimates of the {\it nuclear\/}
X-ray flux of each galaxy.  However, because high-quality HRI profiles are
available for only some of the sources, it would be impossible to apply
corrections to the entire sample in a uniform manner.  On the other hand,
some of the sources are so extended that we can be certain nuclear
components make only minor contributions to their total fluxes.  Several
of these are among the most X-ray--luminous objects in the sample, and
failure to correct for the extended emission would severely distort the
mean AGN-related luminosity.  Thus, for eight objects in Table 3 with
PSPC point-source fractions less than 60\% and high-quality PSPC profiles
(i.e., NGC 2841, 4258, 4472, 4486, 4636, 4736, 5194, and 5195), we extracted 
all counts associated with the central source and adjusted the total nuclear 
X-ray fluxes derived from spectral fitting (discussed in the next section) by 
the PSPC point-source fraction.

The point-source fraction limits obtained for the remainder of the galaxies
are generally too uncertain for this technique to be used reliably.  
Therefore, for all other objects, we have adopted a consistent procedure for 
the determination of nuclear fluxes that is governed by the PSPC
source characteristics, since most of the objects were observed with that
instrument.  Admittedly, because of the modest angular resolution of the
PSPC, nuclear fluxes will be overestimated in some cases.  The radial
profiles of a number of the nuclear sources indicate that the majority of
the extended PSPC emission lies outside the radius that encloses most of the
flux of a point source.  Thus, one way to reduce the contamination of the
nuclear X-ray flux by an extranuclear component would be to extract a
smaller area around the source that contains most of the flux of a point
source and excludes a major fraction of a non-nuclear component.  For each
object detected with the PSPC (except the eight very extended sources
discussed above), we used the theoretical PSFs to determine the radius
that would include 95\% of the counts in a point source.  We then extracted
counts from within that radius, which is approximately $1'$ for on-axis
observations.  (The excluded 5\% of the point-source flux is added back
to the measured flux.)  We also used a local background, starting $1.5'$ 
from the edge of the source region, in order to subtract off some of the 
extended component.  This procedure does not remove all of the effects of 
extended emission; but unlike the point-source fractions derived above, 
which are upper limits whose quality varies considerably from case to case, 
this procedure works uniformly well for all sources regardless of source 
distance, off-axis angle, and detection strength.

\subsection{Spectral Analysis and Luminosity Measurements}

Based on the quality of the {\sl ROSAT\/} detections, we selected a sample
of objects for spectral analysis with {\tt XSPEC}.  We found that a spectrum
with at least 300 net PSPC source counts is required for reliable model
fitting.  In preparation for fitting, spectra were grouped to have a minimum
of 25--50 counts per energy bin, and effective-area response files were
generated using the FTOOLS task {\tt pcarf}.  In each case, the PSPC response
matrix appropriate for the gain setting of the detector was used.  Because
the data have limited spectral resolution and signal-to-noise ratio ($S$/$N$), 
they can only be fitted with relatively simple models.  AGNs often have 
power-law X-ray spectra, so we first attempted to fit each spectrum with a 
simple absorbed power-law model.  In a number of cases, this provided a 
satisfactory fit (i.e., a reduced $\chi^2$ of 1.3 or less).  When power-law 
fits were unacceptable, we tried a two-component model consisting of a power 
law plus a Raymond-Smith thermal plasma (Raymond \& Smith 1977) with solar 
abundances.  This type of model was found to be useful for analysis of 
{\sl ASCA\/} spectra of LLAGNs (Ptak et al.\ 1999; Terashima, Ho, \& Ptak 
2000).  In the two-component models, the absorption column density of the 
Raymond-Smith component was fixed to the Galactic value, because we expect 
this emission to arise from an extranuclear source that is not significantly 
absorbed.  Note that seven of the eight highly extended sources discussed in
\S~4.1 require a Raymond-Smith component.  For the power-law component, the 
absorption was constrained to be greater than or equal to the Galactic value.  
For almost all of the strongly detected objects, we achieved a very good fit 
with one of these two models.

We want to stress that our main purpose for modeling the spectra is to
determine accurate fluxes and luminosities of the galaxies in the sample.
While it is tempting to try to use the spectral parameters associated with
the best-fit models for quantitative comparisons, high-$S$/$N$ X-ray
observations of AGNs with good energy resolution invariably reveal complex
spectral features, and the possibility of contamination by extranuclear
emission in these low-luminosity sources leaves us very suspicious of the
correspondence between our best-fit models and reality.  We have decided,
therefore, not to report the fit parameters.  Nonetheless, it should be
noted that the spectral parameters from our fits are largely consistent
with expected values for both high- and low-luminosity AGNs (Walter \& Fink
1993;  Ptak et al.\ 1999).  For power-law components, the photon index
ranges from 1.1 to 4.3 with a typical value of 2.5, and for Raymond-Smith
components, $kT$ ranges from 0.42 keV to 1.1 keV with a typical value of
0.6 keV.  For each of the galaxies whose spectrum we modeled, we list in
Table 4 the Galactic neutral hydrogen column density, the type of model
used, the reduced $\chi^2$ for the fit, the observed flux, and the
luminosity (corrected for Galactic absorption only).

For the weaker PSPC-observed objects (those with fewer than 300 source
counts) and those observed only with the HRI or the {\sl Einstein\/} IPC,
we must adopt a spectral model in order to convert their detected count
rates to fluxes.  Under the assumption that these sources are similar
to the 20 galaxies in our sample whose spectra we have modeled, we can use
the spectral fitting results obtained above to derive an average model
suitable for this purpose.  Unfortunately, as Table 4 indicates, both one-
and two-component models were needed to obtain good fits, which complicates
the procedure.  However, we have identified a simple one-component (power-law)
model that allows us to estimate source fluxes reliably regardless of the
intrinsic form of the spectrum.  The power-law components used in the models
for the objects listed in Table 4 have a typical photon index of 2.5 and a
column density slightly higher than the Galactic column.  Therefore, we have
selected as a typical model a power law with a photon index $\Gamma$ = 2.5
and column density $N_{\rm H}$ = Galactic + $2 \times 10^{20}$ cm$^{-2}$.  
To test the accuracy of the fluxes predicted by our ``standard'' model,
we applied it to the 20 well-detected objects for which spectral fitting
was possible.  We used the {\tt PIMMS} software to derive, from the standard
model and the observed count rates, fluxes corrected for Galactic absorption
within the region containing 95\% of the PSF counts.  Figure 2 shows a 
comparison of the flux implied by the standard model and that derived from 
the spectral fitting.  The straight line indicates $F$(model) = $F$(fitted).  
As Figure 2 illustrates, the predicted and observed fluxes agree very well 
over the entire range.

Before we apply the standard model to the HRI-only and weak PSPC sources,
we would like to be certain that the spectra of the stronger sources are
indeed representative of the spectra of all the objects in the sample. 
While we cannot produce high-quality spectra for objects with few PSPC 
counts, we can characterize their spectra using a ``hardness ratio.''  The
hardness ratio is defined as ($H-S$)/($H+S$), where $H$ is the number of
counts in a hard energy band and $S$ is the number of counts in a soft band.
We chose 0.7 keV as the cutoff energy between the hard and soft bands and
computed hardness ratios for all objects detected with the PSPC.  Hardness
ratios span a fairly wide range for both the well-detected objects (those
with more 300 PSPC counts) and the weaker sources (those with fewer than 300
counts).  To determine whether the hardness ratio distributions of these
two groups are similar (i.e., whether they could arise from the same
parent distribution), we have compared them using a Kolmogorov-Smirnov test.
The cumulative distribution functions for the hardness ratios of the two
groups of sources are shown in Figure 3.  The maximum value of the 
absolute difference between the functions is $D$ = 0.25.  The probability 
that $D$ would be greater than 0.25 if the data sets were drawn from the 
same distribution is 0.67.  Thus, the two distributions of hardness ratios
are consistent with a single parent distribution.  This indicates that the
spectral properties of the weak sources and those of the strong ones do not
differ significantly, and furthermore, that application of the standard model
to estimate fluxes for the weak-PSPC and HRI-only {\sl ROSAT\/} sources is
appropriate.  These fluxes are also listed in Table 4.  Likewise, for the
two objects observed by {\sl Einstein}, the 0.2--3.5 keV IPC count rates
from Fabbiano et al.\ (1992) were converted to 0.1--2.4 keV fluxes using
the same model.

Uncertainties in the absorption-corrected fluxes depend upon the method by 
which they were obtained.  For the fluxes measured directly from the spectra,
the flux uncertainty depends mainly upon the quality of the fit to the
spectrum.  Typical uncertainties for these fluxes are $\pm 10$\% or less.
Fluxes of the remaining objects were based on both the count rate and use of
the standard spectral model, each of which has associated uncertainties.
The uncertainty in the observed counts $N$ (i.e., $N^{-{1 \over 2}}$) ranges
from about 6\% to about 14\%, depending on the object.  For the uncertainty
associated with the application of the standard model, recall that we tested
the model by applying it to all the bright objects with high-quality spectra
(Fig.~2).  Assuming the distribution of $F$(model)/$F$(fitted) for these
sources is Gaussian, we measure a root-mean-square ({\it rms}) deviation of 
0.11 from the mean ratio.  For a given source, then, the 68\% confidence 
interval on the flux predicted by the standard model is approximately 
$\pm 11$\%.  The uncertainties associated with the count rate and use of the 
standard model are independent, so we may add them in quadrature to estimate 
the overall uncertainty in the flux and luminosity.  In the worst cases, the 
flux uncertainty for these sources is about 20\%. 

\subsection{$L_{\rm X}$ vs.\ $L_{{\rm H}\alpha}$: Predicting X-ray Fluxes of 
Unobserved Galaxies}

A correlation has been established between the soft X-ray luminosities and
\Ha\ luminosities of AGNs (Elvis, Soltan, \& Keel 1984; Koratkar et al.\ 1995).
In principle, a similar relationship amongst the LLAGNs in our sample would 
permit us to estimate X-ray fluxes for those members of the sample that have 
not been observed with {\sl ROSAT\/} or {\sl Einstein}.  This is desirable 
for calculation of the total X-ray volume emissivity represented by this 
sample, which factors into our estimates of the LLAGN contribution to the 
soft X-ray background (\S~5.3).

In Figure 4, we have plotted the X-ray luminosities for the objects
listed in Table 4 against their extinction-corrected \Ha\ luminosities from
Table 1.  The plot includes objects with upper limits on $L_{\rm X}$ and
$L_{{\rm H}\alpha}$.  The typical 10\% uncertainty in $L_{\rm X}$ for the 
objects with fitted spectra is approximately equivalent to the size of the 
symbols plotted, and a typical uncertainty of about 16\% for the 
weaker PSPC sources is shown in the upper left corner of Figure 4.  For most 
objects, the uncertainty in $L_{{\rm H}\alpha}$ is between 10\% and 
30\%---exceptions are described in the Notes to Table 1.  An uncertainty
of 30\% in $L_{{\rm H}\alpha}$ is also indicated in the upper left corner 
of Figure 4.  A comparison of the typical error bars to the distribution of 
the points in the Figure suggests that the scatter in the plot is real, 
and not solely due to errors in our fluxes.  Our results are in good 
agreement with those of Roberts \& Warwick (2000), who investigated the 
$L_{\rm X}$--$L_{{\rm H}\alpha}$ correlation in their sample of galaxies 
(consisting of a wide variety of classifications) observed with the 
{\sl ROSAT\/} HRI.

Despite the scatter, we find a correlation between the soft X-ray and \Ha\ 
luminosities of the LLAGNs in our sample.  The generalized Kendall's $\tau$ 
correlation test (Isobe, Feigelson, \& Nelson 1986), which properly accounts 
for the presence of censored data, yields $\tau = 0.56$ and $Z = 3.01$, 
corresponding to a probability of 0.003 that the two luminosities are 
uncorrelated.  This result is not a spurious distance effect.  To investigate 
the consequence of the mutual dependence of $L_{\rm X}$ and $L_{{\rm H}\alpha}$
on distance, we applied the partial Kendall's $\tau$ test (Akritas \& Siebert 
1996) using distance as the third variable.  The partial Kendall's $\tau$ 
coefficient is 0.29 and the square root of the variance is 0.087 (see 
Akritas \& Siebert), which imply that the null hypothesis of no correlation 
between $L_{\rm X}$ and $L_{{\rm H}\alpha}$ can be rejected at a significance 
level of 0.001.  In terms of a Gaussian probability distribution, this is 
equivalent to $\sim 3$~$\sigma$.  Thus, we feel that the correspondence 
between the two luminosities is significant enough to permit use of the 
relationship for predicting X-ray luminosities of the 14 galaxies in our 
sample lacking {\sl ROSAT\/} data.  

The individual $L_{\rm X}$/$L_{{\rm H}\alpha}$ ratios for the sources 
with high-quality luminosity measurements range from about 1 to 100.  This
is similar to the range found by Koratkar et al.\ at higher luminosities 
($10^{40} < L_{\rm X} < 10^{46}$ ergs~s$^{-1}$).  The median 
$L_{\rm X}$/$L_{{\rm H}\alpha}$ ratio for our sample of LLAGNs is 7, which 
is somewhat close to the average ratio of 14 found for the five least 
luminous sources ($\sim 10^{40}-10^{41}$ ergs s$^{-1}$) in the Koratkar 
et al.\ (1995) study.  We use this median ratio, represented by the line in 
Figure 4, to estimate X-ray luminosities for the 14 galaxies in Table 1 that 
lack soft X-ray data.  The distributions of $L_{\rm X}$/$L_{{\rm H}\alpha}$ 
for the Seyferts and LINERs are virtually identical (we return to this point 
below), so we may apply the conversion independent of the source 
classification.

\subsection{Previous Flux Measurements}

Several of the objects in our sample are well-known galaxies for which
{\sl ROSAT\/} observations have already been published.  We have employed
a uniform procedure for the determination of fluxes and luminosities of
the galaxies, which affords consistent results for our large sample.
But by its nature, such an approach can be insensitive to particular details
associated with specific objects.  Thus, it is useful to compare the
fluxes and luminosities we have derived with those obtained previously as
a result of detailed analyses of the PSPC data.

The fluxes that other investigators have reported are for the 0.1--2.4 keV 
band and are corrected for Galactic absorption only, except as otherwise
noted.  Note that {\sl ROSAT\/} data for other well-known galaxies (such 
as NGC 1068 and NGC 4374) have been published, but fluxes and spectral 
modeling have not been reported.

{\it NGC 4051.}  Komossa \& Fink (1997) used a warm absorber model to derive 
an X-ray flux of $4.0 \times 10^{-11}$ ergs~cm$^{-2}$~s$^{-1}$.  While this
is higher than the flux we measured, they also used a different PSPC 
observation, in which the count rate of this highly variable object was 
greater than the count rate for the observation we selected.

{\it NGC 4258.}  A number of sources in the vicinity of this object were 
investigated by Vogler \& Pietsch (1999).  In an attempt to isolate nuclear 
counts, they extract counts from a very small central region (of radius 
$24''$, corresponding to the FWHM of the on-axis PSPC PSF at 1.0 keV), and 
subtract off a local background.  Their best-fit model for the spectrum of 
the nuclear region consists of a thermal bremsstrahlung with $kT$ = 0.63 keV.  
Their resulting flux is very close to the nuclear flux that we estimated by 
lowering the total flux by the fraction included in a theoretical point 
source (see \S~4.1).

{\it NGC 4278.}  Koratkar et al.\ (1995) fitted the spectrum with an absorbed 
power-law model.  They report an unabsorbed 0.2--2.2 keV flux of 
$2.56 \times 10^{-12}$ ergs~cm$^{-2}$~s$^{-1}$.  Note that this is corrected 
not only for Galactic absorption, but also for the total best-fit column 
density, which exceeds the Galactic value.  If we adopt the same waveband 
and correct for the same column density, we obtain about half their flux.  
Their measured count rate is not listed, but since we used spectral parameters 
very similar to theirs, they presumably extracted counts from a bigger region 
than we did.

{\it NGC 4388.}  Antonelli, Matt, \& Piro (1997) fitted a simple absorbed 
power law to obtain a flux of $6 \times 10^{-13}$ ergs~cm$^{-2}$~s$^{-1}$, 
which is consistent with our value within the expected errors.  However, they 
propose that a Raymond-Smith model with low metal abundance provides a better 
description of the spectrum.  

{\it NGC 4395.}  Using an absorbed power-law model, Moran et al.\ 
(1999) found a mean 0.2--2.0 keV flux of about twice our value, even
though they measured approximately the same count rate from the same data. 
Their best-fit model indicated a much flatter power law ($\Gamma = 0.9$) than 
that which we have adopted here.  Thus, in this case, different assumptions 
about the spectrum have led to different fluxes.  

{\it NGC 4450.}  Komossa, Bohringer, \& Huchra (1999) found, as we did, that 
a single power law with $N_{\rm H}$ close to the Galactic value provides an 
excellent fit to the soft X-ray spectrum.  From this model, they derived a 
flux consistent with our own.

{\it NGC 4486.}  Prieto (1996) attempted to determine the nuclear flux by 
subtracting off a local background and fitting the ``residual'' nuclear 
spectrum with a power law.  The resulting upper limit is about twice our 
value.  We have used a different strategy for isolating the nuclear flux, as 
described in \S~4.1, and expect that our value represents a tighter
constraint on the output of the nucleus.

{\it NGC 4565.}  Optically thin thermal plasma and power-law models were 
tested by Vogler, Pietsch, \& Kahabka (1996) on various sources associated 
with this object.  For the central source, both models lead to fluxes 
consistent with ours.

{\it NGC 4636.}  Trinchieri et al.\ (1994) fitted the spectrum separately in 
concentric annuli around the nucleus.  Using the $10'-15'$ annulus as a local
background, they obtain a 0.1--2.0 keV flux of 
$9.38 \times 10^{-12}$ ergs~cm$^{-2}$~s$^{-1}$ from a central region of 
radius $8'$.  This is comparable to the flux that we derived for the entire
extended source, but we substantially reduced this flux by including only
the fraction contained within a point source.  

{\it NGC 4639.}  Koratkar et al.\ (1995) fitted the spectrum with an absorbed 
power law of photon index 2.27.  The resulting unabsorbed 0.2--2.2 keV flux 
is somewhat higher than ours, but it has been corrected for a best-fit column 
density that exceeds the Galactic $N_{\rm H}$.  If we use the same column
density in our model, we obtain a similar flux.

{\it NGC 4736.}  A two-component model similar to ours, consisting of a power 
law and a Raymond-Smith plasma, was favored by Cui, Feldkhun, \& Braun (1997).
Significant deviation at the low-energy end of the spectrum necessitated the 
addition of a Gaussian line feature centered at 0.22 keV.  Altering elemental 
abundances and experimenting with more complicated models failed to improve 
the low-energy fit, so they suggested that the Gaussian component may be a 
calibration artifact.  They derived an observed 0.1--2.0 keV flux of 
$1.78 \times 10^{-12}$ ergs~cm$^{-2}$~s$^{-1}$, which is very close to our
observed flux before we attempted to minimize the included flux due to 
extended emission.  Again, we have attempted to estimate only the flux arising 
from a point source, so we expect that our value is closer to the flux 
produced by the nucleus itself.  

As a whole, while some differences exist between our fluxes and those measured
previously, most appear to be due to variability, different extractions of 
counts, or our attempts to eliminate extended emission, rather than 
substantially different spectral models.  Thus, we have confidence in the 
statistical reliability of our relatively uniform approach.  

\section{Discussion}

\subsection{Comparing LINERs and Seyfert Galaxies}

Comparison of the soft X-ray properties of Seyfert galaxies and LINERs may 
help to determine whether they share a similar type of power source.  The
{\sl ROSAT\/} data we have assembled are particularly suited for such a
comparison, since our sample of LLAGNs is volume-limited and complete.
Therefore, we do not expect selection effects to bias our conclusions.

Table 5 lists the median values of X-ray luminosity and 
$L_{\rm X}$/$L_{{\rm H}\alpha}$ for the various subtypes of galaxies 
in the sample.  We also list the interquartile ranges, which contain the 
middle 50\% of the values.  Because these values include limits, we used the 
Kaplan-Meier product-limit estimator (Feigelson \& Nelson 1985) to compute 
the medians and quartiles.  In Figure 5, we compare the distributions of 
$L_{\rm X}$ and $L_{\rm X}$/$L_{{\rm H}\alpha}$ for Seyfert galaxies and 
LINERs; upper limits are indicated by arrows.  Examination of Figure 5 and 
Table 5 reveals that the typical X-ray luminosity of Seyferts is very close 
to that of LINERs, and that the distributions of 
$L_{\rm X}$/$L_{{\rm H}\alpha}$ for Seyferts and LINERs also appear very 
similar.  We have applied statistical tests for censored univariate data in 
order to quantify the apparent similarity.  The Gehan and logrank tests 
(Feigelson \& Nelson 1985) indicate, respectively, 60\% and 98\% probabilities 
that the $L_{\rm X}$ samples arise from the same parent population, and 19\% 
and 22\% probabilities that the $L_{\rm X}$/$L_{{\rm H}\alpha}$ distributions 
are the same.  Because two distributions are not considered statistically 
different unless the null hypothesis can be rejected at less than 5\%, we 
conclude that $L_{\rm X}$ and $L_{\rm X}$/$L_{{\rm H}\alpha}$ are indeed 
statistically similar for our Seyferts and LINERs.   

Although we have not presented the details of the spectral modeling used to 
measure the fluxes of the well-detected LLAGNs, we found no significant 
differences between the soft X-ray spectral characteristics of the Seyferts 
and LINERs.  For example, Seyfert and LINER spectra were equally likely to 
require a Raymond-Smith component in addition to a power law.  The 
luminosity and spectral similarities may therefore be an indication that, 
in general, Seyferts and LINER nuclei are powered by the same physical 
mechanism.  The existence of a correlation between the X-ray and \Ha\ 
fluxes of the galaxies in our sample and the absence of any segregation of 
the Seyferts and LINERs in Figure 4 provides additional support for this 
conclusion.

We note, however, that many of our objects exhibit significant extended
X-ray emission, which complicates the determination of the X-ray flux
associated with the nucleus.  As discussed in \S~4, we have attempted to
minimize the contribution of extended emission in our flux measurements.
Nevertheless, the possibility remains that extranuclear components are
responsible for non-negligible amounts of the measured flux in many 
sources---even those that are approximately point-like---due to the
modest angular resolution of the PSPC.  The lower point-source fractions
obtained from HRI observations of the same sources (see \S~4.1 and Table 3)
suggest this might be the case.  Thus, the similarities of the mean soft
X-ray luminosities and $L_{\rm X}$/$L_{{\rm H}\alpha}$ ratios for Seyfert 
galaxies and LINERs may be due to the presence of significant amounts of 
circumnuclear extended emission in {\it both\/} types of objects.  If so, 
it would be premature to conclude that their power sources are similar.

As a preliminary test of this possibility, we have attempted to measure
objectively the role of extended emission in the objects for which we
constructed HRI radial profiles.  At the distance limit of our sample
(18 Mpc), more than 90\% of the energy in the HRI PSF is contained within a
radius of $\sim 0.2'$, which corresponds to a physical radius of
$\sim$ 1 kpc.  We have therefore defined a parameter $\xi =$ (source counts
detected beyond 1 kpc)/(total source counts) -- $f$, where $f$ is the
fraction of the energy in the PSF found at $r > 1$ kpc (for the most
distant sources, $f$ has a maximum value of $\sim 0.07$).  Thus, $\xi$
will have a value of zero for an unresolved source and a value near
unity for a very extended source; in general, it represents a lower
limit on the fraction of the central X-ray emission produced by
non-nuclear components.  We have measured $\xi$ for all sources with HRI
radial profiles.  The distribution of $\xi$ values for Seyferts and
LINERs is plotted in Figure 6.  The distributions are quite similar,
indicating that the low-luminosity Seyferts and LINERs have, in a 
statistical sense, comparable amounts of extended central X-ray emission.
Of course, this is a very tentative conclusion---HRI profiles are
available for only a fraction of our sample, and their quality
varies considerably from source to source.

In summary, the soft X-ray properties of our LLAGN sample provide ambiguous
evidence regarding the existence of a common excitation mechanism in Seyfert
galaxies and LINERs.  On one hand, we would not expect Seyferts and LINERs 
to occupy the same region in Figure 4 if the majority of LINERs are
powered by stellar processes rather than accretion onto a massive compact
object.  On the other hand, the $L_{\rm X}$-$L_{{\rm H}\alpha}$ correlation 
we have obtained is not very strong, and we have demonstrated clearly that 
a significant fraction of the central X-ray emission in LLAGNs (of all types) 
is extranuclear in origin.  A definitive comparison of the X-ray properties 
of Seyfert galaxies and LINERs will thus require the angular resolution and 
energy range afforded by {\sl Chandra}.

\subsection{Comparing Type 1 and Type 2 AGNs}

The unified model for AGNs holds that the principal difference between
type 1 and type 2 objects is their orientation with respect to the viewer
(e.g., Antonucci 1993).  We observe a type 1 AGN when we look directly into 
the broad-line region, whereas type 2 objects are observed when the broad-line 
region is obscured by an optically thick torus.  If the unified model is 
universal, we would expect type 1 galaxies to have generally higher soft 
X-ray luminosities than type 2s.  Furthermore, because of the additional 
absorbing material along the line of sight to type 2 nuclei, 
$L_{\rm X}$/$L_{{\rm H}\alpha}$ should be significantly greater in type 1 
objects.  

The distributions of these quantities are shown in Table 5.  While 
$L_{\rm X}$ tends to be somewhat higher for type 1s than type 2s (see Table 
7), the Gehan and logrank tests give 5\% and 6\% probabilities that the 
$L_{\rm X}$ samples are the same, indicating that $L_{\rm X}$ may tend to 
be marginally higher for type 1 objects.  For the 
$L_{\rm X}$/$L_{{\rm H}\alpha}$ distributions, however, the tests indicate 
13\% and 11\% probabilities, respectively, that the samples arise from the 
same parent population, so we are unable to conclude that a substantial 
difference exists.  At face value, then, these data do not strongly support 
the general applicability of the unified model.  If we assume that the 
majority of the X-ray luminosity actually arises from an AGN in most cases, 
the similarity of the $L_{\rm X}$ and $L_{\rm X}$/$L_{{\rm H}\alpha}$ 
distributions suggests that type 2 objects are not, in fact, significantly 
more absorbed than type 1 objects.  On the other hand, the spectra of type 2 
objects are somewhat more likely to require a Raymond-Smith component, and 
we have found that many of the galaxies in our sample have substantial 
extended components that may be contaminating their nuclear emission.  Since 
we do not expect the flux produced outside the nucleus to be highly absorbed, 
such contamination may be masking any intrinsic differences between the soft 
X-ray properties of type 1 and type 2 nuclei.

\subsection{Contribution of LLAGNs to the Cosmic X-ray Background}

A substantial fraction of the cosmic X-ray background (XRB) is known to arise
from QSOs and other highly luminous AGNs.  However, the luminosity function,
clustering properties, and X-ray spectral characteristics of such objects
indicate that they cannot account for all of the observed background (Fabian
\& Barcons 1992).  Contributions from  other types of sources, such as non-AGN 
galaxies and/or obscured LLAGNs, may be required to provide a full description 
of the observed XRB (Comastri et al.\ 1995).  In this section, we attempt to
quantify the XRB contribution of low-luminosity Seyferts and LINERs such as
those in our sample.

First, we must compute the local soft X-ray volume emissivity of LLAGNs.  
Our unbiased sample allows us to estimate this quantity accurately.  To 
avoid the possibility of incompleteness in the ``zone of avoidance'' we 
have only included galaxies that are more than $30^{\circ}$ from the Galactic 
plane (this is 53 of the objects in our sample).  In this region, the total 
0.1--2.4 keV luminosity of the objects that were observed by {\sl ROSAT\/} 
is $1.70 \times 10^{42}$ ergs~s$^{-1}$ (uncorrected for absorption).  The 
sources lacking {\sl ROSAT\/} data (see \S~4.3) contribute an estimated 
additional $1.16 \times 10^{41}$ ergs~s$^{-1}$, for a total energy output of 
$1.81 \times 10^{42}$ ergs~s$^{-1}$ from LINERs and Seyferts.  Assuming a 
power-law spectrum with $\Gamma = 2.5$ and 
$N_{\rm H}$ = $4 \times 10^{20}$ cm$^{-2}$, this translates to an 
absorption-corrected volume emissivity of
$2.20 \times 10^{38}$ ergs~s$^{-1}$~Mpc$^{-3}$ in the 0.5--2.0 keV band.
This result is in close agreement with the value of 
$2.27 \times 10^{38}$ ergs~s$^{-1}$~Mpc$^{-3}$ obtained
for low-luminosity X-ray galaxies via optical/X-ray cross-correlation
techniques (Almaini et al.\ 1997).

Assuming that the LLAGN volume emissivity does not evolve with redshift, we
have computed the XRB intensity produced by such objects by integrating eq.\
(18) from Soltan et al.\ (1996):  $$I_{\rm X} = {{c \over {4 \pi H_0}}\, \rho \, }
{\int\displaylimits_{0}^{z_{\rm max}}}
{{(1 + z)^{1 - \alpha}} \over {(1 + z)^3\, \sqrt{1 + \Omega_{\rm M} z}}\, }
dz$$ where $\rho$ is the volume emissivity, $\Omega_{\rm M}$ is the density 
parameter, $z$ is the redshift, and $\alpha$ ($= \Gamma - 1$) is the average 
energy index of their power-law X-ray spectra.  We chose $z = 5$ as the upper 
limit of integration; the calculation is not sensitive to this choice, since 
more than 95\% of the XRB contribution is produced at $z < 2$ in the absence 
of evolution.  We assume the total 0.5--2 keV XRB intensity is 
$2.61 \times 10^{-8}$ ergs~cm$^{-2}$~s$^{-1}$~sr$^{-1}$ (Soltan et al.\ 1996). 
In an $\Omega_{\rm M} = 1.0$, $\Omega_{\Lambda} = 0.0$ cosmology, we find that 
nonevolving LLAGNs produce 9\% of the soft
XRB.  Generalizing the Soltan et al.\ equation for an $\Omega_{\rm M} = 0.3$,
$\Omega_{\Lambda} = 0.7$ cosmology (e.g., Garnavich et al. 1998; Balbi et al.
2000; Hanany et al. 2000), we find their contribution is 11\%.  A higher 
fraction of the soft XRB will be accounted for by LLAGNs if their volume 
emissivity evolves with redshift.  We note, however, that if LLAGNs were to 
evolve as strongly as high-luminosity AGNs (see Miyaji, Hasinger, \& Schmidt 
2000), they would {\it overproduce\/} the soft XRB.  This supports the 
conclusion of Miyaji et al.\ that there is a strong luminosity dependence
on the evolution of the AGN X-ray luminosity function.

To compare our results with recent estimates of the soft XRB contribution
from higher luminosity AGNs, we consider the findings from the {\sl ROSAT\/}
Deep Survey of the ``Lockman Hole.''  Hasinger et al.\ (1998) resolved 
68\%--81\% of the 1--2 keV X-ray background in this field, depending on the 
actual level and spectrum of the XRB in that energy range.  Subsequent optical
follow-up by Schmidt et al.\ (1998) revealed that 39 of the 50 sources
resolved in the survey are luminous AGNs (most having luminosities between
$10^{43}$ and $10^{45}$ ergs~s$^{-1}$), indicating that 53\%--63\% of the
XRB is produced by luminous AGNs in the 1--2 keV band.  In comparison,
the values of 9\%--11\% we estimate for the LLAGN contribution to the soft
XRB, which are lower limits, appear to be reasonable.

\section{Summary}

We have analyzed the soft X-ray properties of a complete, distance-limited
sample of 60 galaxies classified by Ho et al.\ (1997a) as Seyfert galaxies 
or LINERs.  We find that the central X-ray sources in the majority of the 
galaxies exhibit significant amounts of extended emission, in addition to 
the emission from the active nucleus.  We have attempted to isolate the 
nuclear emission as much as possible, guided by spatial information contained 
in the X-ray images.  The spectra of objects with at least 300 net PSPC counts 
are well fitted with either simple absorbed power-law models, or with 
two-component models consisting of a power law and a thermal plasma.  For 
these objects, fluxes and luminosities were derived from the spectral 
modeling; a standard spectral model based on their typical fit parameters was 
used to determine the fluxes of the weaker sources.  We have investigated the 
relationship between $L_{\rm X}$ and $L_{{\rm H}\alpha}$ for the 46 objects 
with soft X-ray data, finding a weak correlation between these quantities 
that is roughly consistent with the correlation found in higher-luminosity 
AGNs.  We have used the median $L_{\rm X}$/$L_{{\rm H}\alpha}$ ratio of 7
for our sample to predict the X-ray luminosities of the 14 objects lacking
soft X-ray fluxes.

We find that low-luminosity Seyferts and LINERs have similar soft X-ray
properties.  In terms of their X-ray luminosity, 
$L_{\rm X}$/$L_{{\rm H}\alpha}$ distribution, and spectral properties, 
no differences are apparent, suggesting that the Seyferts and the majority 
of LINERs may be powered by a common mechanism.  It is possible, however, 
that a general presence of significant extended emission in both types of 
objects is responsible for the apparent similarities.  A comparison of the 
X-ray properties of the type 1 and type 2 AGNs in our sample did not reveal 
the differences expected if the unified AGN model is universal, although 
once again, contamination of the nuclear X-ray fluxes by extended emission 
components could be responsible.

Finally, we have estimated the fraction of the soft X-ray background
contributed by low-luminosity Seyferts and LINERs, based on the local
0.5--2.0 keV volume emissivity of
$2.2 \times 10^{38}$ ergs~s$^{-1}$~Mpc$^{-3}$ derived for our sample
of LLAGNs.  With no evolution, our results suggest that LLAGNs produce
9\%--11\% of the XRB in this energy range.

\acknowledgments
This research has made use of data obtained from the High Energy Astrophysics
Science Archive Research Center (HEASARC), provided by NASA's Goddard Space
Flight Center.  The work of ECM is supported by NASA through Chandra
Fellowship PF8-10004 awarded by the Chandra X-ray Observatory Center, which
is operated by the Smithsonian Astrophysical Observatory for NASA under
contract NAS8-39073.  We also acknowledge the support of NASA through grants
NAG 5-3556 and STScI GO-8607.  Finally, we would like to thank the anonymous 
referee for providing many valuable suggestions.

\begin{center}
\begin{deluxetable}{llcl}
\tablewidth{0pt}
\tablecaption{The Distance-Limited LLAGN Sample}
\tablehead{\colhead{~~~~~~~~Galaxy~~~~~~~~} &
           \colhead{Class\tablenotemark{a}~~~~~} &
           \colhead{~~~~~~~$D$ (Mpc)~~~~~~~} &
           \colhead{log $L_{{\rm H}\alpha}$\tablenotemark{b}~~~~~~}}
\startdata
IC 239\tablenotemark{c}   & L2::  & 16.8 & 36.89x \\
NGC 185   & S2    & 0.7 & 34.90  \\     
NGC 404   & L2    & 2.4 & 37.76  \\
NGC 428\tablenotemark{c}  & L2:   & 14.9 & 36.98n \\   
NGC 1052  & L1.9  & 17.8 & 39.80x \\         
NGC 1058  & S2    & 9.1 & 38.16  \\          
NGC 1068  & S1.9  & 14.4 & 41.65x \\       
NGC 2681  & L1.9  & 13.3 & 39.27n \\      
NGC 2683\tablenotemark{c} & L2    & 5.7 & 37.48  \\
NGC 2685\tablenotemark{c} & S2:   & 16.2 & 39.21  \\
NGC 2787\tablenotemark{c} & L1.9  & 13.0 & 38.95  \\ 
NGC 2841  & L2    & 12.0 & 38.80x \\        
NGC 3031  & S1.5  & 1.4 & 38.53  \\          
NGC 3368\tablenotemark{c} & L2    & 8.1 & 38.91x \\
NGC 3379  & L2::  & 8.1 & 37.94c \\
NGC 3486  & S2    & 7.4 & 37.85  \\
NGC 3623  & L2:   & 7.3 & 37.77  \\        
NGC 3718\tablenotemark{c} & L1.9  & 17.0 & 39.29n \\ 
NGC 3982  & S1.9  & 17.0 & 39.82  \\ 
NGC 4051  & S1.2  & 17.0 & 40.47x \\       
NGC 4111\tablenotemark{c} & L2    & 17.0 & 39.84  \\
NGC 4138  & S1.9  & 17.0 & 38.99  \\       
NGC 4143\tablenotemark{c} & L1.9  & 17.0 & 39.27  \\ 
NGC 4168  & S1.9: & 16.8 & 38.40  \\         
NGC 4203  & L1.9  & 9.7 & 38.79  \\        
NGC 4258  & S1.9  & 6.8 & 39.08  \\         
NGC 4278  & L1.9  & 9.7 & 39.38  \\        
NGC 4293  & L2    & 17.0 & 39.68  \\        
NGC 4314  & L2    & 9.7 & 38.56  \\        
\tablebreak
\tablebreak
\tablebreak
\tablebreak
NGC 4346  & L2::  & 17.0 & 37.53c \\         
NGC 4374  & L2    & 16.8 & 39.31  \\        
NGC 4388  & S1.9  & 16.8 & 40.28x \\      
NGC 4394  & L2    & 16.8 & 38.68  \\        
NGC 4395  & S1.8  & 3.6 & 38.67  \\               
NGC 4438  & L1.9  & 16.8 & 40.11  \\        
NGC 4450  & L1.9  & 16.8 & 38.79x \\          
NGC 4457\tablenotemark{c} & L2    & 17.4 & 39.77  \\
NGC 4472  & S2::  & 16.8 & 37.59c \\          
NGC 4477  & S2    & 16.8 & 39.06  \\        
NGC 4486  & L2    & 16.8 & 39.77  \\        
NGC 4494  & L2::  & 9.7 & 37.54u \\        
NGC 4501  & S2    & 16.8 & 39.06  \\  
NGC 4548  & L2    & 16.8 & 38.89  \\
NGC 4550  & L2    & 16.8 & 38.51x \\            
NGC 4565  & S1.9  & 9.7 & 38.70  \\           
NGC 4579  & S1.9  & 16.8 & 39.72  \\         
NGC 4596\tablenotemark{c} & L2::  & 16.8 & 37.95b \\
NGC 4636  & L1.9  & 17.0 & 38.63  \\        
NGC 4639  & S1.0  & 16.8 & 39.77  \\        
NGC 4651  & L2    & 16.8 & 38.24  \\        
NGC 4698  & S2    & 16.8 & 38.74  \\        
NGC 4725  & S2:   & 12.4 & 38.22  \\        
NGC 4736  & L2    & 4.3 & 37.81x \\         
NGC 4762  & L2:   & 16.8 & 37.49b \\         
NGC 4772  & L1.9  & 16.3 & 38.95n \\         
NGC 4866\tablenotemark{c} & L2    & 16.0 & 38.64  \\
NGC 5194  & S2    & 7.7  & 39.80x \\          
NGC 5195  & L2:   & 9.3  & 38.67x \\           
NGC 7217\tablenotemark{c} & L2    & 16.0 & 39.67  \\
\tablebreak
\tablebreak
\tablebreak
\tablebreak
NGC 7814\tablenotemark{c} & L2::  & 15.1 & 36.83b 
\enddata
\tablecomments{Data from Ho et al.\ (1997a).}
\tablenotetext{a}{L = LINER, S = Seyfert; type 2 objects have no detectable 
broad lines, whereas type 1.9 objects exhibit weak broad \Ha\ emission; 
luminosity includes broad and narrow components where both exist.  Uncertain 
classifications are followed by a colon; double colons mean that the 
classification is highly uncertain.}
\tablenotetext{b}{``x'' denotes data from one of the following sources:
Stauffer (1982), Keel (1983), or Heckman, Balick, \& Crane (1980).  ``u'' 
indicates a 3~$\sigma$ upper limit, and ``b'' and ``c'' are quality 
ratings corresponding to probable uncertainties of $\pm 30$\%--50\% and 
$\pm 100$\%, respectively.  ``n'' indicates data obtained under 
nonphotometric conditions, which lead to typical uncertainties of 
$\pm 100$\%.}
\tablenotetext{c}{Soft X-ray data are not available.}
\end{deluxetable}
\end{center}

\clearpage
\begin{center}
\begin{deluxetable}{lccccc}
\tablewidth{0pt}
\tablecaption{Soft X-ray Observations} 
\tablehead{\colhead{Galaxy} &
  	   \colhead{Instrument} &
	   \colhead{Count rate\tablenotemark{a}} &
 	   \colhead{Exp.\ (s)} &
	   \colhead{Off-axis angle ($'$)} &
           \colhead{$\Delta_{\rm OX}('')$}}
\startdata
NGC 185  & HRI   & $<$ 1.53E$-$3 &  21070  &  0.9 & \nodata \\
\\
NGC 404  & HRI   &     (1.93 $\pm$ 0.28)E$-$3 &  23874  &  0.2 &  4      \\
\\
NGC 1052 & HRI   &     (8.64 $\pm$ 0.63)E$-$3 &  22105  &  4.4 &  4      \\ 
         & PSPC  &     (3.60 $\pm$ 0.16)E$-$2 &  13975  &  0.4 &  7      \\
\\
NGC 1058 & HRI   & $<$ 9.81E$-$4 &  60796  &  1.3 & \nodata \\
\\
NGC 1068 & HRI   &     (5.80 $\pm$ 0.02)E$-$1 & 114768  &  0.2 &  2      \\ 
         & PSPC  &     1.85 $\pm$ 0.02      &   5471  &  0.2 &  6      \\
\\
NGC 2681 & PSPC  & $<$ 8.78E$-$3 &   4648  & 35.3 & \nodata \\
\\
NGC 2841 & PSPC  &     (4.71 $\pm$ 0.19)E$-$2 &  13428  &  0.2 &  2      \\
\\
NGC 3031 & HRI   &     (2.83 $\pm$ 0.02)E$-$1 & 102000  &  0.3 &  7      \\  
         & PSPC  &     (7.98 $\pm$ 0.04)E$-$1 &  49197  &  0.3 &  7      \\ 
\\
NGC 3379 & HRI   &     (3.22 $\pm$ 0.36)E$-$3 &  24560  &  0.4 &  4      \\
\\
NGC 3486 & HRI   & $<$ 1.69E$-$3 &  15841  &  0.8 & \nodata \\
\\
NGC 3623 & HRI   &     (3.74 $\pm$ 0.32)E$-$3 &  36916  &  0.2 &  8      \\
         & PSPC  &     (1.90 $\pm$ 0.10)E$-$2 &  17494  &  0.2 &  7      \\
\\
NGC 3982 & PSPC  &     (1.45 $\pm$ 0.15)E$-$2 &   6264  &  0.4 &  3      \\
\\
NGC 4051 & HRI   &     1.08 $\pm$ 0.01     &  10579  &  0.1 &  7      \\ 
         & PSPC  &     1.58 $\pm$ 0.01     &  28727  &  1.2 &  9      \\ 
\\
NGC 4138 & HRI   & $<$ 3.43E$-$3 &   5798  &  3.2 & \nodata \\
\\
NGC 4168 & PSPC  &     (7.97 $\pm$ 0.75)E$-$3 &  14300  & 31.1 & 17      \\
\\
NGC 4203 & HRI   &     (5.39 $\pm$ 0.14)E$-$2 &  25448  &  1.0 &  5      \\
         & PSPC  &     (2.34 $\pm$ 0.03)E$-$1 &  22663  &  1.0 &  5      \\
\\
NGC 4258 & HRI   &     (8.74 $\pm$ 0.18)E$-$2 &  27556  &  0.4 & 14      \\
         & PSPC  &     (2.96 $\pm$ 0.03)E$-$1 &  32864  &  0.3 & 15      \\ 
\\
NGC 4278 & HRI   &     (1.64 $\pm$ 0.13)E$-$2 &   9986  &  0.1 &  4      \\ 
         & PSPC  &     (5.45 $\pm$ 0.40)E$-$2 &   3411  &  0.1 & 10      \\
\\
NGC 4293 & PSPC  &     (6.11 $\pm$ 0.59)E$-$3 &  17507  &  0.7 & 11      \\ 
\\
NGC 4314 & PSPC  &     (1.20 $\pm$ 0.10)E$-$2 &  10956  & 22.9 &  7      \\
\\
NGC 4346 & PSPC  & $<$ 1.83E$-$2 &   5510  & 42.3 & \nodata \\
\\
NGC 4374 & HRI   &     (4.80 $\pm$ 0.13)E$-$2 &  26490  &  0.3 &  3      \\
         & PSPC  &     (1.38 $\pm$ 0.03)E$-$1 &  22020  & 17.1 &  9      \\
\\
NGC 4388 & HRI   &     (1.26 $\pm$ 0.11)E$-$2 &  11274  &  0.5 &  3      \\
         & PSPC  &     3.67E$-$2 &  11639  &  1.1 &  9      \\
\\
NGC 4394 & PSPC  & $<$ 3.08E$-$3 &   8495  &  7.2 & \nodata \\
\\
NGC 4395 & HRI   & $<$ 2.49E$-$3 &  11353  &  0.5 & \nodata \\
         & PSPC  &     (8.22 $\pm$ 0.69)E$-$3 &  17038  &  1.2 &  5      \\
\\
NGC 4438 & HRI   &     (1.21 $\pm$ 0.07)E$-$2 &  21651  &  2.0 &  4      \\
         & PSPC  &     (6.10 $\pm$ 0.23)E$-$2 &  11639  & 35.6 & 19      \\
\\
NGC 4450 & PSPC  &     (9.32 $\pm$ 0.25)E$-$2 &  15307  &  1.7 &  4      \\ 
\\
NGC 4472 & HRI   &     (1.84 $\pm$ 0.02)E$-$1 &  34423  &  0.6 &  4      \\ 
         & PSPC  &     (2.37 $\pm$ 0.03)E$-$1 &  25951  &  1.0 &  7      \\
\\
NGC 4477 & PSPC  &     (3.74 $\pm$ 0.23)E$-$2 &   7351  & 10.3 & 15      \\
\\
NGC 4486 & HRI   &     1.50 $\pm$ 0.01       &  45108  &  0.3 &  3      \\ 
         & PSPC	 &    11.86 $\pm$ 0.02      &  30435  &  0.3 &  4      \\
\\
NGC 4494 & PSPC  &     (2.34 $\pm$ 0.14)E$-$2 &  12015  & 45.3 & 13      \\
\\
NGC 4501 & HRI   &     (1.16 $\pm$ 0.11)E$-$2 &  10418  &  5.6 & 14      \\
\\
NGC 4548 & PSPC  & $<$ 2.01E$-$2 &   1262  & 44.7 & \nodata \\
\\
NGC 4550 & PSPC  &     (1.43 $\pm$ 0.09)E$-$2 &  16660  & 15.0 & 10      \\
\\
NGC 4565 & HRI   &     (1.05 $\pm$ 0.14)E$-$2 &   5312  &  1.4 & 12      \\
         & PSPC  &     (2.41 $\pm$ 0.11)E$-$2 &  19707  &  1.1 &  5      \\
\\
NGC 4579 & PSPC  &     (6.36 $\pm$ 0.08)E$-$1 &   9313  &  0.7 &  2      \\
\\
NGC 4636 & HRI   &     (1.94 $\pm$ 0.03)E$-$1 &  22050  &  0.2 & 10      \\ 
         & PSPC  &     (8.54 $\pm$ 0.08)E$-$1 &  13070  &  0.2 & 10      \\
\\
NGC 4639 & HRI   &     (2.08 $\pm$ 0.16)E$-$2 &   8083  &  0.5 &  4      \\
         & PSPC  &     (3.47 $\pm$ 0.23)E$-$2 &   6604  &  0.5 &  7      \\
\\
NGC 4651 & HRI   &     (1.50 $\pm$ 0.24)E$-$3 &  25396  &  2.4 &  9      \\
         & PSPC  &     (1.28 $\pm$ 0.11)E$-$2 &  10481  &  1.2 &  9      \\
\\
NGC 4698 & IPC   &     (7.05 $\pm$ 0.83)E$-$3 &  10355  &  0.2 & \nodata \\
\\
NGC 4725 & PSPC  &     (6.10 $\pm$ 0.18)E$-$2 &  19362  & 14.6 &  9      \\ 
\\
NGC 4736 & HRI   &     (9.05 $\pm$ 0.09)E$-$2 & 112910  &  0.6 &  3      \\
         & PSPC  &     (2.53 $\pm$ 0.02)E$-$1 &  94819  &  0.2 &  1      \\
\\
NGC 4762 & IPC   &     (1.41 $\pm$ 0.11)E$-$2 &  12243  &  5.3 & \nodata \\
\\
NGC 4772 & HRI   & $<$ 2.23E$-$3 &  12120  &  0.6 & \nodata \\
\\
NGC 5194 & HRI   &     (3.83 $\pm$ 0.09)E$-$2 &  45715  &  0.5 &  5      \\ 
         & PSPC  &     (2.24 $\pm$ 0.03)E$-$1 &  23956  &  0.5 &  9      \\
\\
NGC 5195 & HRI   &     (9.19 $\pm$ 0.45)E$-$3 &  45715  &  4.4 &  9      \\ 
         & PSPC  &     (3.31 $\pm$ 0.12)E$-$2 &  23956  &  4.4 & 12      
\enddata
\tablenotetext{a}{Count rates have been corrected for vignetting.  Upper 
limits are 3~$\sigma$.}
\end{deluxetable}
\end{center}

\clearpage
\begin{center}
\begin{deluxetable}{lcc}
\tablewidth{0pt}
\tablecaption{Point-Source Flux Fractions}
\tablehead{\colhead{} &
	   \multicolumn{2}{c}{$\Sigma$(PSF)/$\Sigma$(source)}\\
           \cline{2-3}\\
           \colhead{Galaxy} &
           \colhead{PSPC} &
           \colhead{HRI}}
\startdata
NGC 1052 &   0.88 &   0.34 \\
NGC 1068 &   0.79 &   0.38 \\
NGC 2841 &   0.50 &   \nodata \\
NGC 3031 &   0.78 &   0.79 \\
NGC 3379 &\nodata &   0.47 \\
NGC 3623 &   0.68 &   0.82 \\
NGC 3982 &   1.00 &   \nodata \\
NGC 4051 &   0.83 &   0.56 \\
NGC 4203 &   0.88 &   0.59 \\
NGC 4258 &   0.10 &   0.02 \\
NGC 4278 &   1.00 &   0.61 \\
NGC 4293 &   1.00 &   \nodata \\
NGC 4374 &   0.76 &   0.11 \\
NGC 4388 &   0.61 &   0.24 \\
NGC 4395 &   1.00 &   \nodata \\
NGC 4438 &\nodata &   0.31 \\
NGC 4450 &   0.91 &   \nodata \\
NGC 4472 &   0.11 &   0.03 \\
NGC 4477 &   0.86 &   \nodata \\
NGC 4486 &   0.06 &   0.03 \\
NGC 4501 &\nodata &   0.26 \\  
NGC 4550 &   1.00 &   \nodata \\ 
NGC 4565 &   1.00 &   1.00 \\
NGC 4579 &   1.00 &   \nodata \\
NGC 4636 &   0.13 &   0.03 \\
NGC 4639 &   1.00 &   0.70 \\
NGC 4651 &   1.00 &   0.82 \\
NGC 4725 &   0.83 &   \nodata \\
NGC 4736 &   0.47 &   0.11 \\
NGC 5194 &   0.33 &   0.16 \\
NGC 5195 &   0.25 &   0.13 
\enddata
\end{deluxetable}
\end{center}

\clearpage
\begin{center}
\begin{deluxetable}{lccccc}
\tablewidth{0pt}
\tablecaption{Observed X-ray Fluxes and Luminosities}
\tablehead{\colhead{Galaxy} &
           \colhead{$N_{\rm H}$\tablenotemark{a}} &
           \colhead{Model} &
           \colhead{$\chi_{\nu}^{2}$} &
	   \colhead{$F_{\rm X}$\tablenotemark{b}} &
           \colhead{$\log L_{\rm X}$\tablenotemark{c}}}
\startdata
NGC 185  & 12.3 & \nodata  & \nodata &$<$ 0.534 &$<$ 36.83  \\
NGC 404  & 5.31 & \nodata  & \nodata &    0.684 &    37.87  \\
NGC 1052 & 3.07 & PL       & 1.03    &    4.439 &    40.32  \\
NGC 1058 & 6.65 & \nodata  & \nodata &$<$ 0.347 &$<$ 38.77  \\
NGC 1068 & 3.53 & PL+RS    & 1.18    &  171.3   &    41.90  \\
NGC 2681 & 2.45 & \nodata  & \nodata &$<$ 1.056 &$<$ 39.53  \\
NGC 2841\tablenotemark{d} & 1.45 & PL       & 1.25    &    1.970 &    39.64  \\
NGC 3031 & 4.16 & PL       & 1.27    &  103.1   &    39.50  \\
NGC 3379 & 2.75 & \nodata  & \nodata &    1.121 &    39.07  \\
NGC 3486 & 1.91 & \nodata  & \nodata &$<$ 0.607 &$<$ 38.67  \\
NGC 3623 & 2.16 & \nodata  & \nodata &    1.807 &    39.16  \\
NGC 3982 & 1.22 & \nodata  & \nodata &    1.528 &    39.79  \\
NGC 4051 & 1.32 & PL+RS    & 1.14    &  115.5   &    41.72  \\
NGC 4138 & 1.36 & \nodata  & \nodata &$<$ 1.153 &$<$ 39.67  \\ 
NGC 4168 & 2.56 & \nodata  & \nodata &    0.823 &    39.55  \\
NGC 4203 & 1.19 & PL       & 1.28    &   21.25  &    40.45  \\
NGC 4258\tablenotemark{d} & 1.16 & PL+RS    & 1.07    &    2.501 &    39.22  \\
NGC 4278 & 1.77 & \nodata  & \nodata &    5.921 &    39.91  \\
NGC 4293 & 2.58 & \nodata  & \nodata &    0.673 &    39.48  \\
NGC 4314 & 1.78 & \nodata  & \nodata &    1.355 &    39.27  \\
NGC 4346 & 1.13 & \nodata  & \nodata &$<$ 1.904 &$<$ 39.88  \\
NGC 4374 & 2.60 & PL+RS    & 1.28    &   13.39  &    40.76  \\
NGC 4388 & 2.60 & PL+RS    & 1.25    &    3.949 &    40.24  \\
NGC 4394 & 2.52 & \nodata  & \nodata &$<$ 0.344 &$<$ 39.18  \\
NGC 4395 & 1.35 & \nodata  & \nodata &    0.862 &    38.20  \\
NGC 4438 & 2.66 & PL       & 1.18    &    7.430 &    40.53  \\
NGC 4450 & 2.37 & PL       & 1.10    &    9.839 &    40.65  \\
NGC 4472\tablenotemark{d} & 1.66 & PL+RS    & 1.08    &    8.179 &    40.56  \\
NGC 4477 & 2.64 & \nodata  & \nodata &    4.237 &    40.28  \\
NGC 4486\tablenotemark{d} & 2.54 & PL+RS    & 1.39    &   72.12  &    41.58  \\
NGC 4494 & 1.52 & \nodata  & \nodata &    2.564 &    39.54  \\
NGC 4501 & 2.48 & \nodata  & \nodata &    4.019 &    40.25  \\
NGC 4548 & 2.36 & \nodata  & \nodata &$<$ 2.354 &$<$ 40.01  \\
NGC 4550 & 2.57 & \nodata  & \nodata &    1.654 &    39.87  \\
NGC 4565 & 1.30 & PL       & 0.94    &    3.224 &    39.57  \\
NGC 4579 & 2.47 & PL       & 1.05    &   78.79  &    41.51  \\
NGC 4636\tablenotemark{d} & 1.81 & PL+RS    & 1.01    &   10.42  &    40.66  \\
NGC 4639 & 2.35 & \nodata  & \nodata &    4.104 &    40.25  \\
NGC 4651 & 1.99 & \nodata  & \nodata &    1.200 &    39.70  \\
NGC 4698 & 1.87 & \nodata  & \nodata &    1.599 &    39.83  \\
NGC 4725 & 1.00 & PL       & 1.00    &    3.264 &    39.91  \\
NGC 4736\tablenotemark{d} & 1.44 & PL+RS    & 1.11    &    9.095 &    39.49  \\
NGC 4762 & 2.04 & \nodata  & \nodata &    3.382 &    40.14  \\
NGC 4772 & 1.79 & \nodata  & \nodata &$<$ 0.798 &$<$ 39.47  \\
NGC 5194\tablenotemark{d} & 1.57 & PL+RS    & 0.96    &    6.339 &    39.76  \\ 
NGC 5195\tablenotemark{d} & 1.56 & PL+RS    & 1.35    &    0.845 &    39.00  
\enddata
\tablecomments{If the flux was obtained from a fitted spectrum, the type of 
spectral model and the reduced $\chi^2$ for the best fit are given, and 
typical errors are $\pm 10$\% or less.  Otherwise, the flux was obtained with 
the count rate and the ``standard'' model, with uncertainties of $\pm 20$\% 
or less; see \S~4.2.}
\tablenotetext{a}{From Dickey \& Lockman (1990); listed in units of 
$10^{20}$ cm$^{-2}$.}
\tablenotetext{b}{0.1--2.4 keV observed fluxes in units of 
$10^{-13}$ erg s$^{-1}$ cm$^{-2}$.}
\tablenotetext{c}{0.1--2.4 keV luminosities (ergs s$^{-1}$) have been 
corrected for Galactic absorption.}
\tablenotetext{d}{One of the eight very extended objects which received the 
special correction for extended emission discussed in \S~4.1.}
\end{deluxetable}
\end{center}

\clearpage
\begin{center}
\begin{deluxetable}{lcccc}
\tablewidth{0pt}
\tablecaption{X-ray Luminosity Statistics}
\tablehead{\colhead{} &
	   \multicolumn{2}{c}{$\log (L_{\rm X}$/ergs~s$^{-1}$)} &
           \multicolumn{2}{c}{$L_{\rm X}$/$L_{{\rm H}\alpha}$} \\
           \cline{2-3} \cline{4-5}\\
	   \colhead{} &
           \colhead{Median} &
           \colhead{IQR\tablenotemark{a}} &
           \colhead{Median} &
           \colhead{IQR\tablenotemark{a}}}
\startdata
All objects & 39.66 & 39.14--40.27 & ~7.0 & 1.4--27.8 \\
LINERs      & 39.54 & 39.11--40.28 & ~7.0 & 2.1--45.1 \\
Seyferts    & 39.76 & 38.88--40.25 & ~3.0 & 0.8--15.5 \\
Type 1      & 39.91 & 39.36--40.53 & ~3.1 & 0.9--14.1 \\
Type 2      & 39.52 & 39.01--39.89 & 13.4 & 1.4--28.7 
\enddata
\tablenotetext{a}{IQR = interquartile ratio; the range containing the middle
50\% of the values.}
\end{deluxetable}
\end{center}

\clearpage
\figcaption[]{($a$) PSPC and ($b$) HRI radial profiles ({\it crosses}) of the
nuclear X-ray sources in several galaxies, along with the theoretical PSFs
({\it solid lines}) corresponding to the source position on the relevant
detector.  Below each profile, the normalized residuals (i.e., [source
profile -- PSF] / [$1~\sigma$ source profile uncertainty]) are plotted.
To the right of the profiles are the {\sl ROSAT\/} images of the sources,
which are $12'$ square in ($a$) and $6'$ square in ($b$).  Coordinates are
J2000.  The nucleus is centered in these images.  The different plots 
represent the range of profiles observed in our LLAGN sample: point-like 
(NGC 4450 and NGC 4565), slightly extended (NGC 4725 and NGC 4051), and 
very extended (NGC 4258 and NGC 4736).}

\begin{figure}
\vskip -0.92truein
\centerline{\psfig{figure=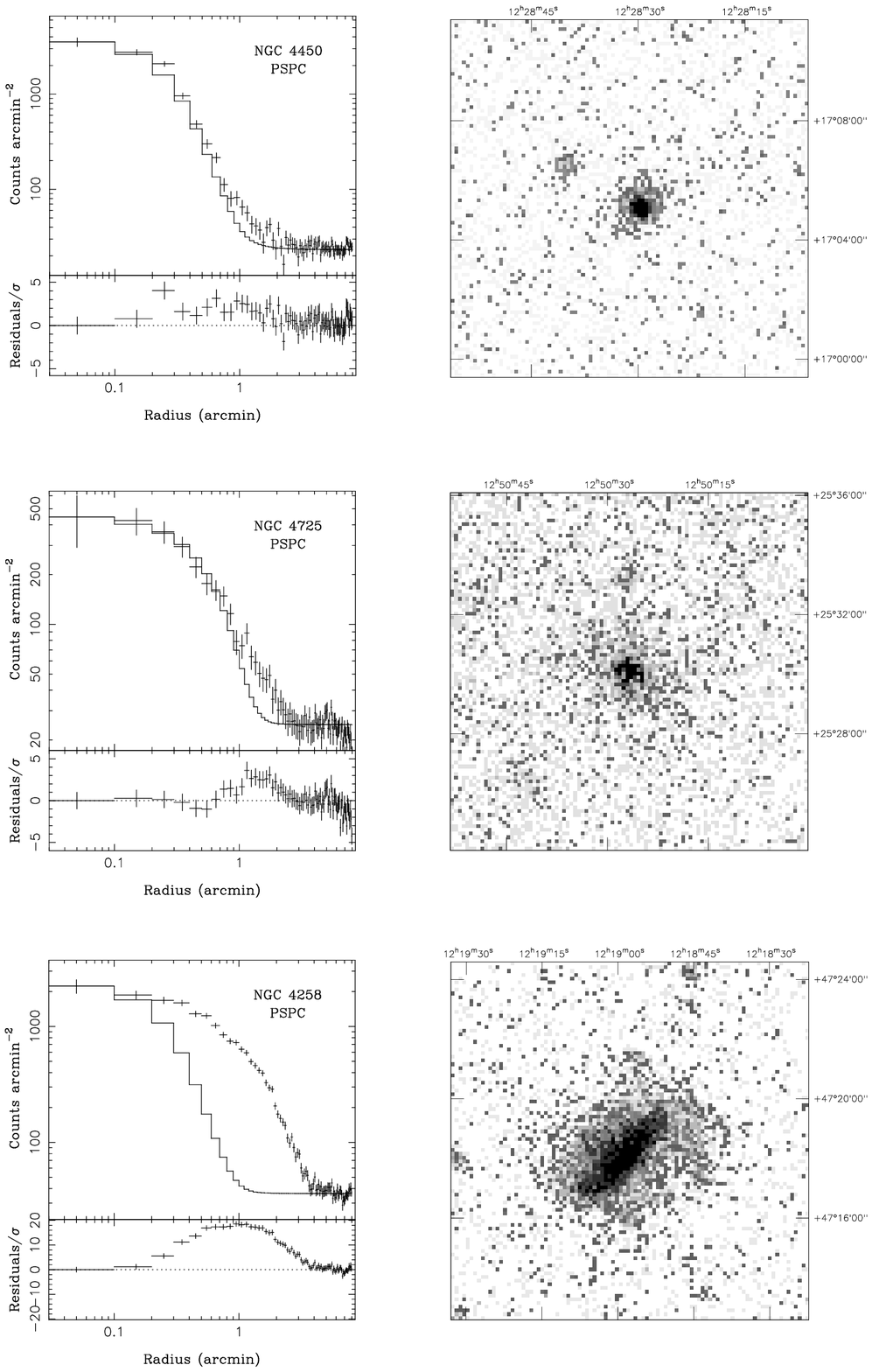,height=10truein,angle=0}}
\vskip -0.6truein
\centerline{Fig.~1a}
\end{figure}

\begin{figure}
\vskip -0.92truein
\centerline{\psfig{figure=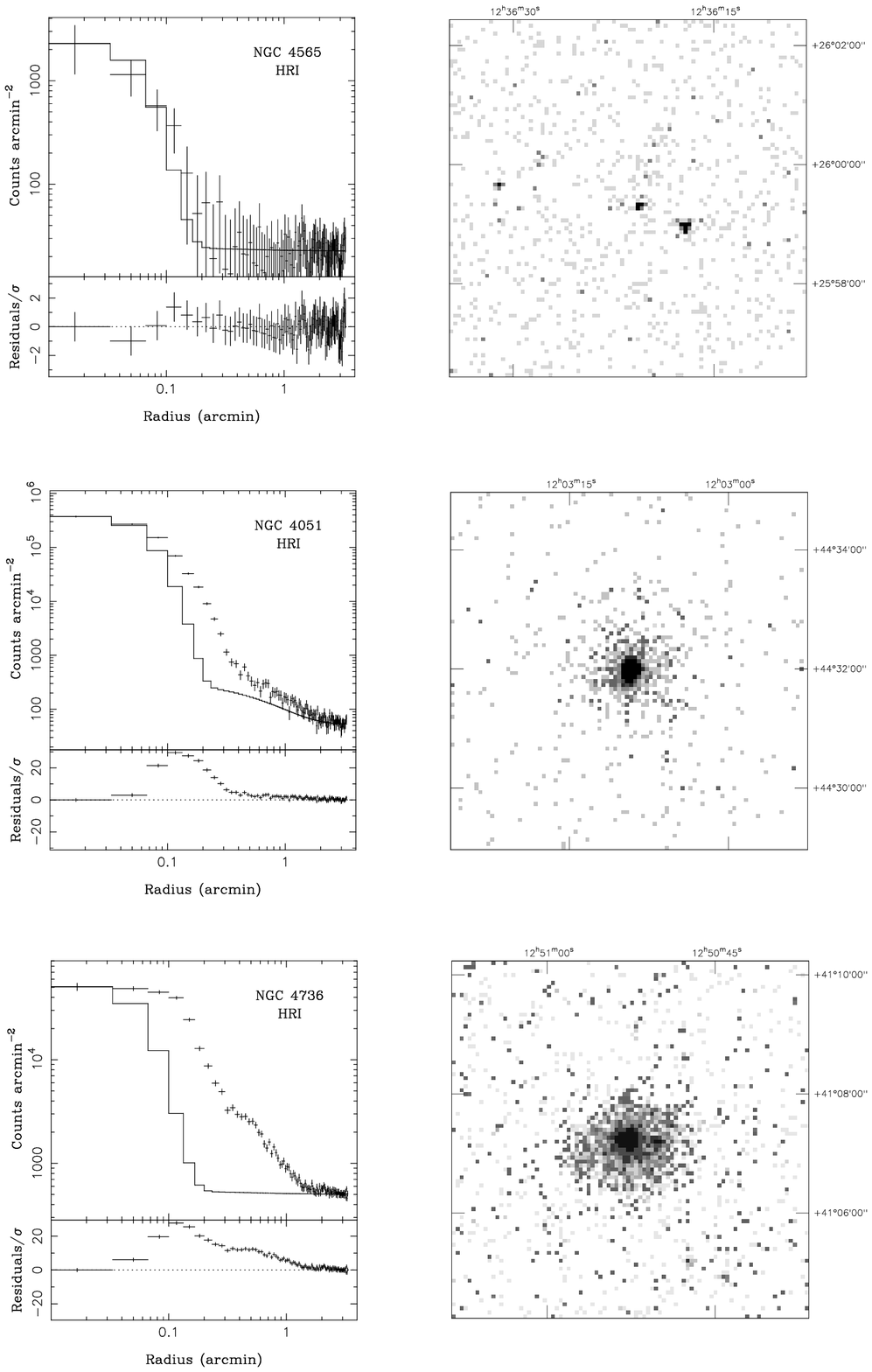,height=10truein,angle=0}}
\vskip -0.6truein
\centerline{Fig.~1b}
\end{figure}

\begin{figure}
\begin{center}
\centerline{\psfig{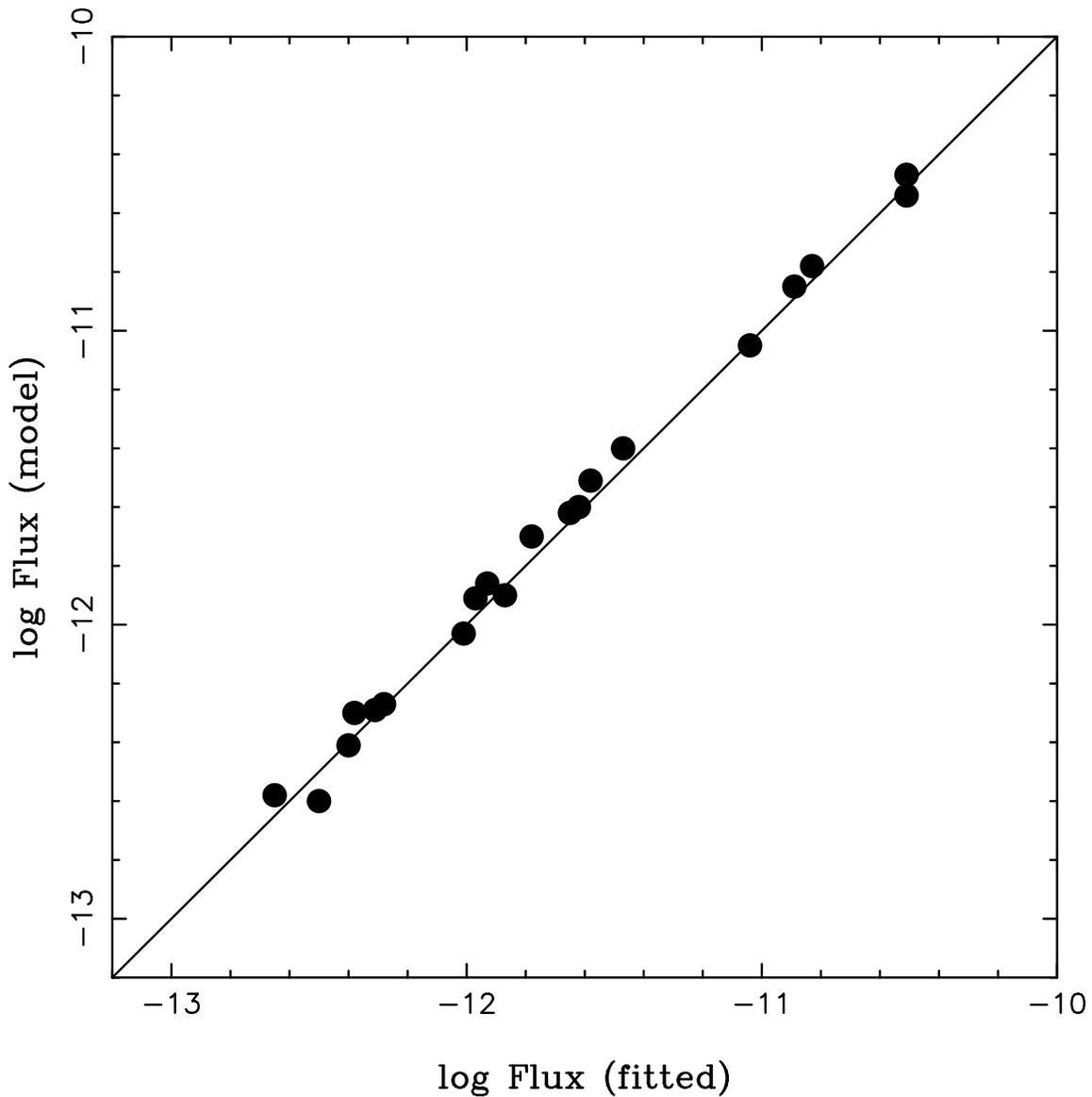}}
\vskip 0.2truein
\caption{Comparison of 0.1--2.4 keV fluxes derived using the observed
count rate and the ``standard'' spectral model with those derived from careful
model fitting, for well-detected PSPC sources (i.e., those with at least
300 net source counts).  The solid line indicates $F$(model) = $F$(fitted). 
The close agreement of the two flux determinations indicates that the standard 
model may be applied to obtain accurate fluxes for weaker sources.}
\end{center}
\end{figure}

\begin{figure}
\begin{center}
\centerline{\psfig{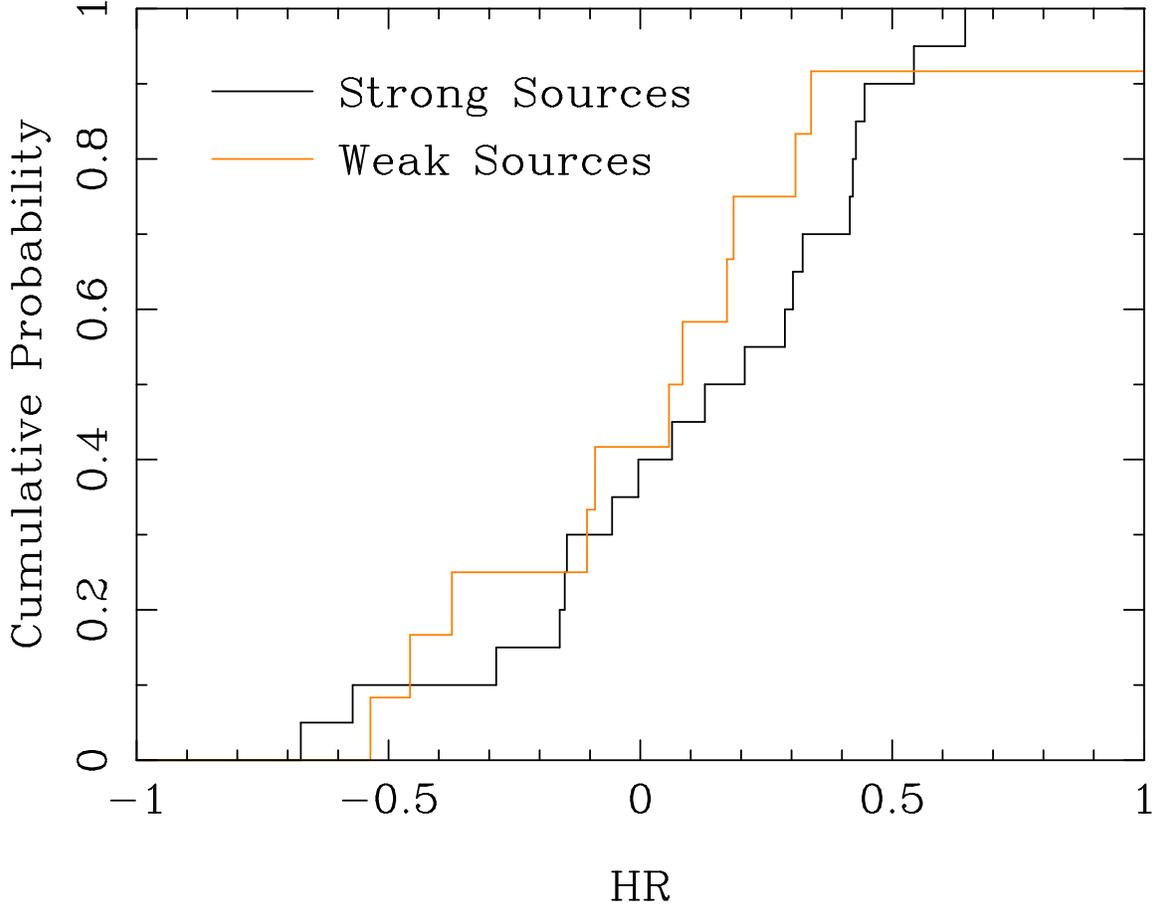}}
\vskip 0.2truein
\caption{Cumulative probability distributions of the hardness ratios
of strongly and weakly detected sources.  The maximum value of the 
absolute difference between the functions is $D$ = 0.25.  If the data sets 
were drawn from the same distribution, the probability that $D$ would be 
higher is 0.67.  Thus, these distributions indicate that the spectra of the 
two types of sources do not differ significantly.}
\end{center}
\end{figure}

\begin{figure}
\begin{center}
\centerline{\psfig{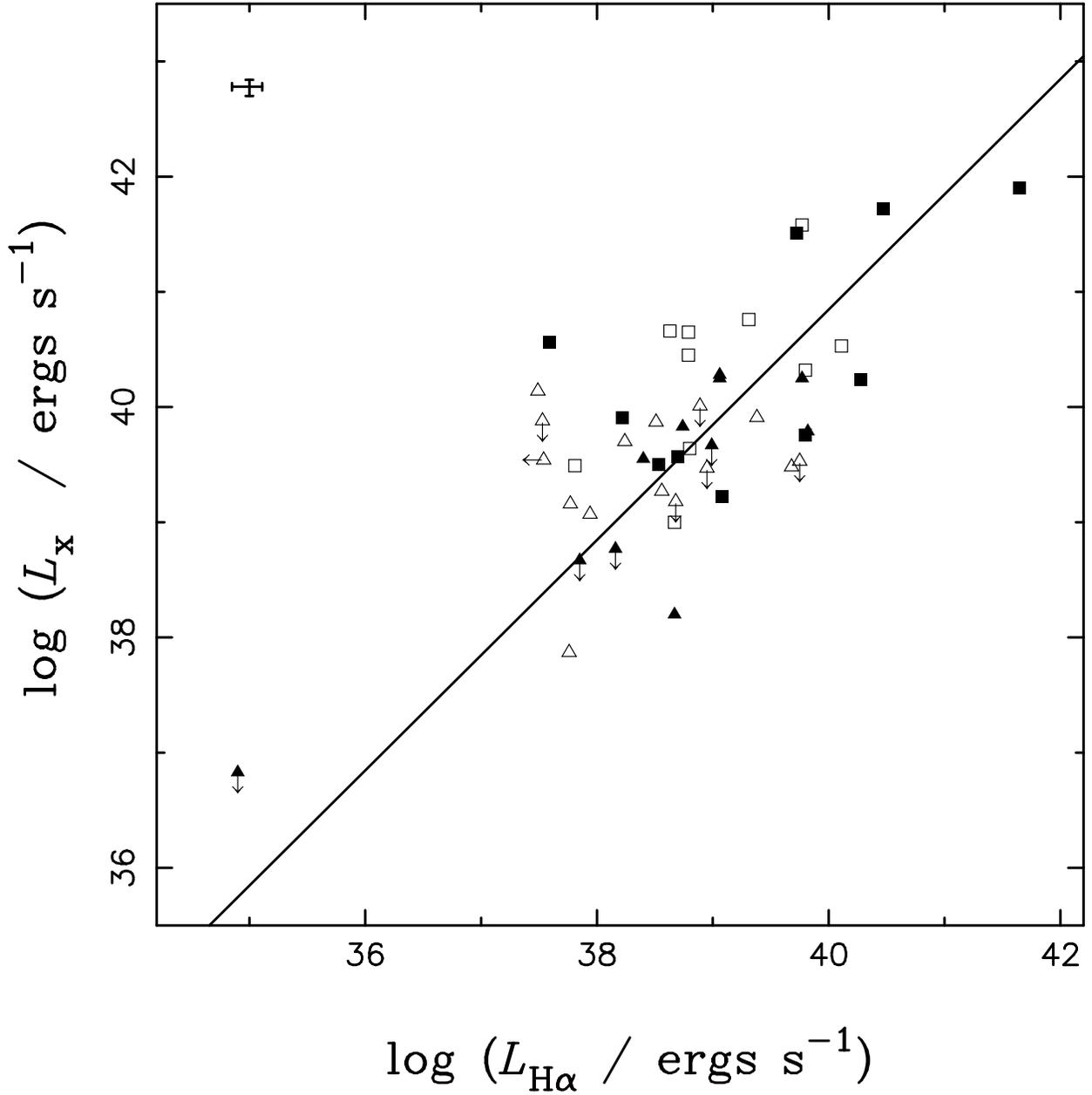}}
\vskip 0.2truein
\caption{$L_{\rm X}$ vs.\ $L_{{\rm H}\alpha}$ for all objects with
soft  X-ray observations.  The line represents the median 
$L_{\rm X}$/$L_{{\rm H}\alpha}$ ratio of 7.  Seyfert galaxies are represented
by  filled symbols, and LINERs by open symbols.  Squares indicate that 
$L_{\rm X}$ was obtained from spectral modeling; triangles indicate that 
$L_{\rm X}$ was derived from the observed count rate and standard spectral 
model.  Upper and lower limits are indicated by arrows.  The error bars in 
the upper left corner represent a typical uncertainty in $L_{\rm X}$ of 
about 16\% for the weaker PSPC sources, and a 30\% uncertainty in 
$L_{{\rm H}\alpha}$.}
\end{center}
\end{figure}

\begin{figure}
\begin{center}
\centerline{\psfig{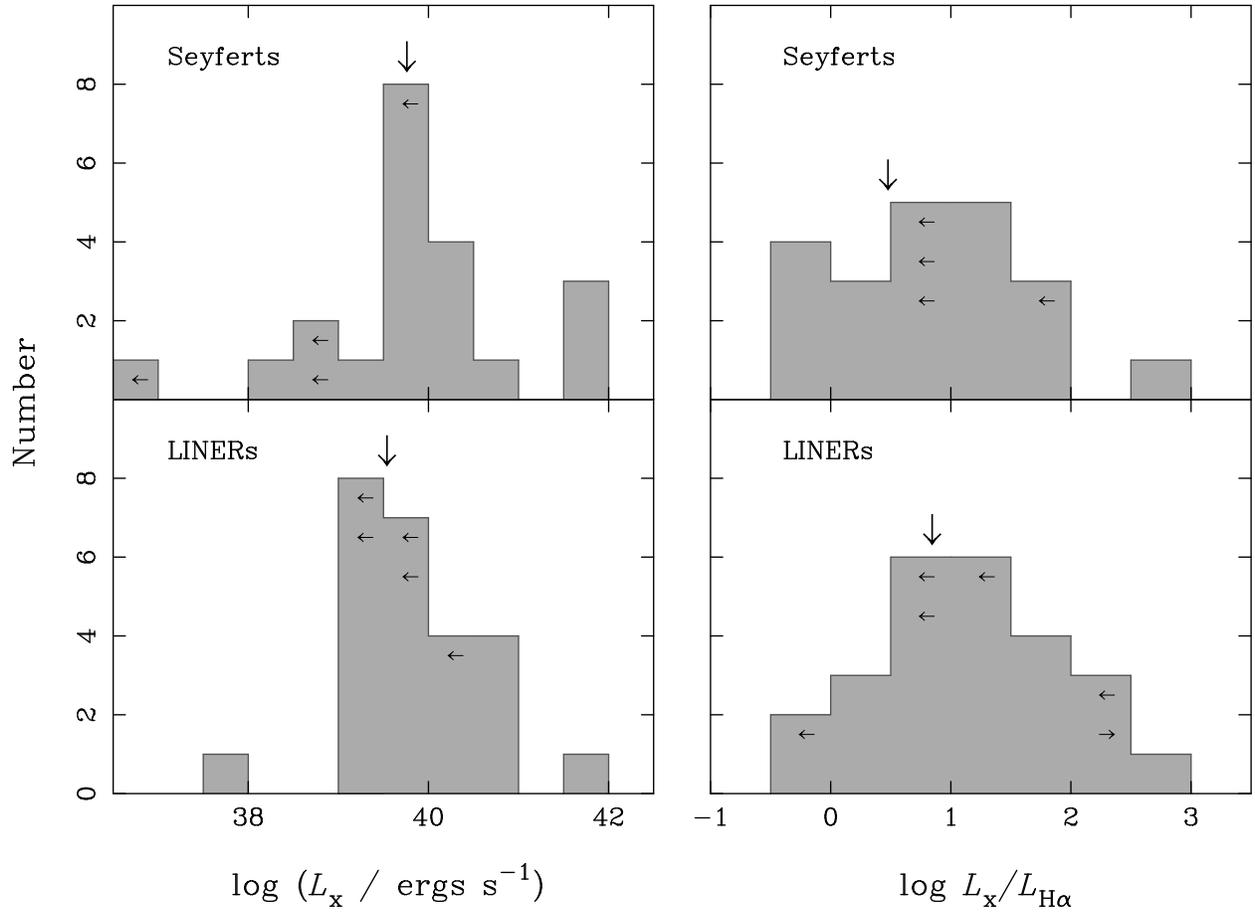}}
\vskip 0.2truein
\caption{Distributions of $L_{\rm X}$ and $L_{\rm X}$/$L_{{\rm H}\alpha}$
for the Seyferts and LINERs in our sample.  Median values are indicated by 
arrows above the histograms.}
\end{center}
\end{figure}

\begin{figure}
\begin{center}
\centerline{\psfig{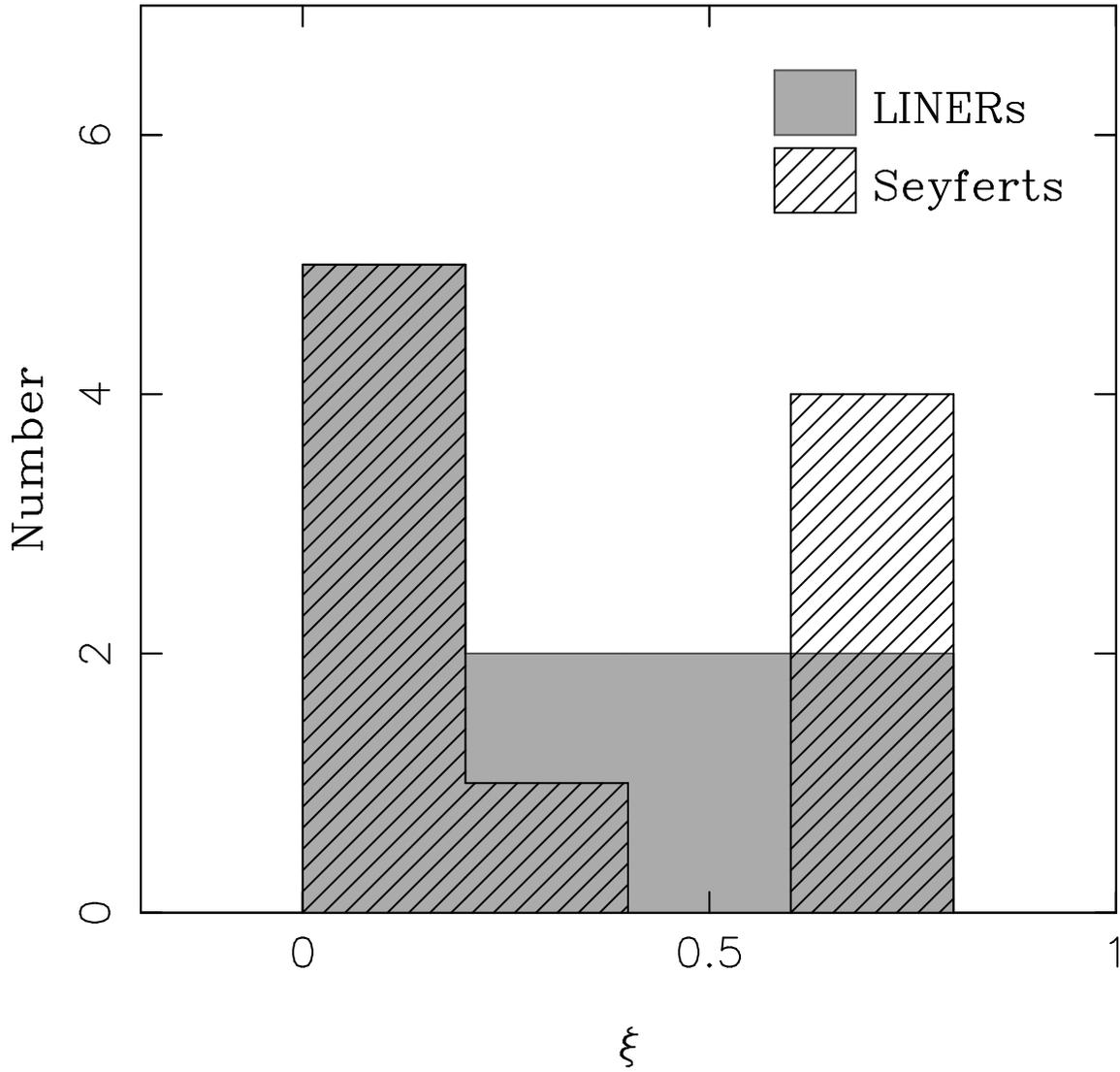}}
\vskip 0.2truein
\caption{Distributions of the extended emission parameter $\xi$ for
Seyferts and LINERs with measured HRI radial profiles.}
\end{center}
\end{figure}

\begin{figure}
\begin{center}
\centerline{\psfig{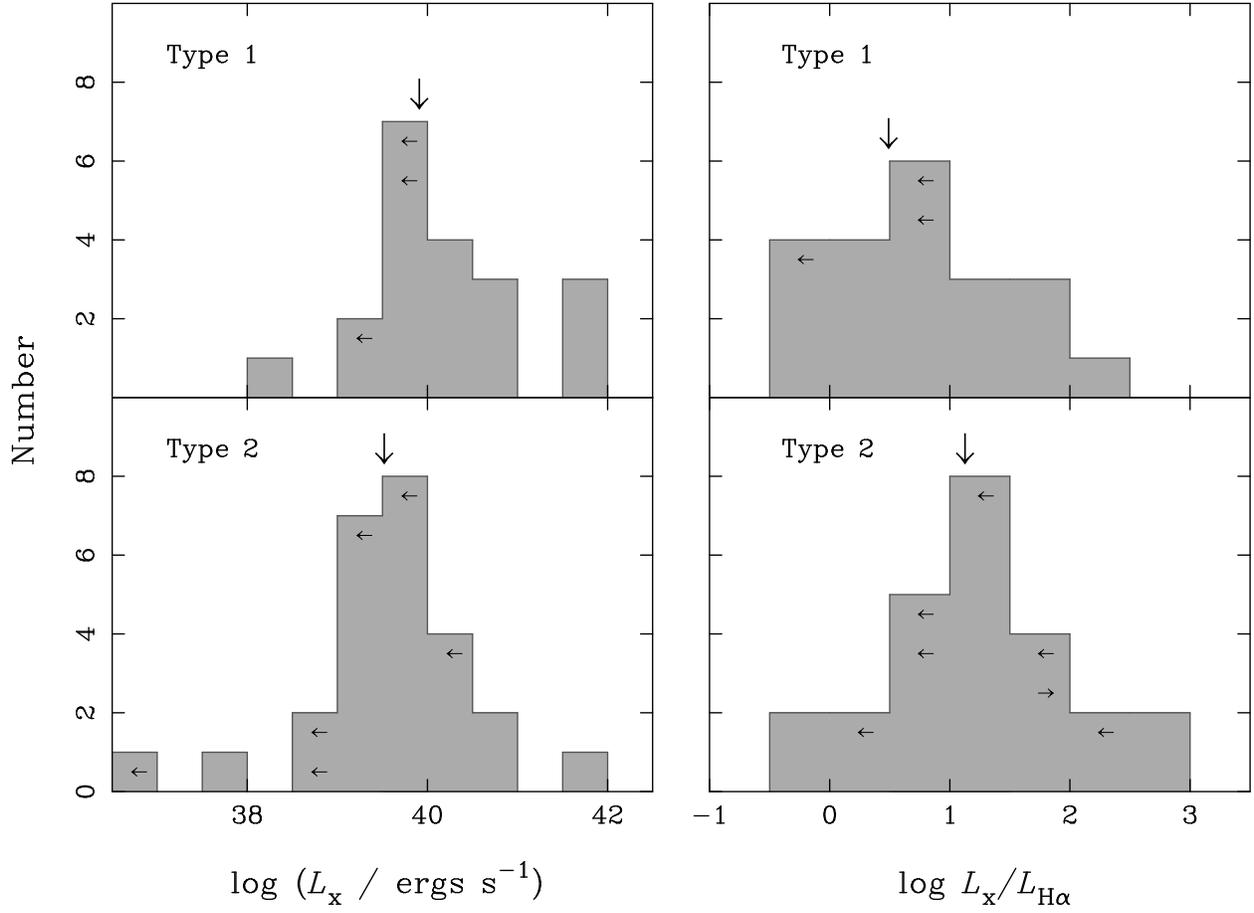}}
\vskip 0.2truein
\caption{Distributions of $L_{\rm X}$ and $L_{\rm X}$/$L_{{\rm H}\alpha}$
for the type 1 and type 2 AGNs in our sample.  Median values are indicated
by arrows above the histograms.}
\end{center}
\end{figure}

\end{document}